\documentclass[sigconf,authorversion,screen]{acmart}

\AtBeginDocument{%
  \providecommand\BibTeX{{%
    \normalfont B\kern-0.5em{\scshape i\kern-0.25em b}\kern-0.8em\TeX}}}

\setcopyright{acmcopyright}
\copyrightyear{2023}
\acmYear{2023}
\acmDOI{10.1145/3576915.3623187}

\acmConference[CCS]{Make sure to enter the correct
  conference title from your rights confirmation emai}{November 26-30, 2023}{Copenhagen, Denmark}
\acmBooktitle{ACM CCS:ACM Conference on Computer and Communications Security,Copenhagen, Denmark}

\usepackage{float}
\usepackage{algorithm}
\usepackage{algorithmic}
\usepackage{amsmath}
\usepackage{subcaption}
\usepackage{enumitem}

\newcommand\system{{\sc ProvG-Searcher}\space}
\newcommand\systemNoSpace{{\sc ProvG-Searcher}}

\newcommand\eat[1]{{}}
\usepackage{multirow}

\begin{document}

\title{\systemNoSpace: A Graph Representation Learning Approach for Efficient Provenance Graph Search}

\author{Enes Altinisik}
\email{ealtinisik@hbku.edu.qa}
\affiliation{%
\institution{Qatar Computing Research Institute}
\country{HBKU, Qatar}
}
\author{Fatih Deniz}
\email{fdeniz@hbku.edu.qa}
\affiliation{%
\institution{Qatar Computing Research Institute}
\country{HBKU, Qatar}
}
\author{Husrev Taha Sencar}
\email{hsencar@hbku.edu.qa}
\affiliation{%
\institution{Qatar Computing Research Institute}
\country{HBKU, Qatar}
}

\renewcommand{\shortauthors}{Altinisik, et al.}
\begin{abstract}

\textcolor{black}{We present \systemNoSpace, a novel approach for detecting known APT behaviors within system security logs.
Our approach leverages provenance graphs, a comprehensive graph representation of event logs, to capture and depict data provenance relations by mapping system entities as nodes and their interactions as edges.
We formulate the task of searching provenance graphs as a subgraph matching problem and employ a graph representation learning method.
The central component of our search methodology involves embedding of subgraphs in a vector space where subgraph relationships can be directly evaluated. 
We achieve this through the use of order embeddings that simplify subgraph matching to straightforward comparisons between a query and precomputed subgraph representations.
To address challenges posed by the size and complexity of provenance graphs, we propose a graph partitioning scheme and a behavior-preserving graph reduction method.  
Overall, our technique offers significant computational efficiency, allowing most of the search computation to be performed offline while incorporating a lightweight comparison step during query execution. 
Experimental results on standard datasets demonstrate that \system achieves superior performance, with an accuracy exceeding 99\% in detecting query behaviors and a false positive rate of approximately 0.02\%, outperforming other approaches. 
}

\end{abstract}


\maketitle

\renewcommand{\headrulewidth}{0.0pt}
\thispagestyle{fancy}
\lhead{}
\rhead{}
\chead{To Appear in 2023 ACM Conference on Computer and Communications Security (CCS), November 2023}
\cfoot{}

\section{Introduction}

Causality analysis and provenance graphs have emerged as crucial tools for understanding and mitigating risks associated with cyber attacks targeting computer systems.
A provenance graph is a holistic representation of kernel audit logs, describing interactions between system entities \textcolor{black}{\cite{moreau2008open, muniswamy2010provenance}}.
By interconnecting isolated system events, provenance graphs offer two indispensable capabilities for security analysis.
First, they facilitate identifying causal relationships and tracing data lineage, thereby providing an enriched context for discerning the nature of system events.
Second, they enable the application of advanced graph algorithms in audit log analysis. 
Consequently, provenance graph analysis has been extensively employed in the  detection of anomalous system activities 
\cite{manzoor2016fast,liu2019log2vec, wang2020you, han2020unicorn,li2021hierarchical, wang2022threatrace,zengy2022shadewatcher,fang2022back}; 
root-cause analysis and forensic tracking \cite{king2003backtracking,liu2018towards,hossain2018dependence,hassan2020we,fei2021seal};
attack story generation \cite{pei2016hercule, hossain2017sleuth, hossain2020combating, alsaheel2021atlas}; and 
supporting alert validation and investigation \cite{hassan2019nodoze, milajerdi2019holmes,xiong2020conan, hassan2020tactical}.

Another domain that can greatly benefit from using provenance graphs is the efficient search for known attack behaviors within vast repositories of historical system logs. 
This is an under-studied problem with crucial implications for the practice of threat hunting. 
Due to the increasing complexity and volume of cyber attacks, relying solely on detection tools available within organizations is no longer feasible for in-house security teams. 
Consequently, threat hunters must continuously scan descriptions of new threat behaviors provided by threat intelligence sources and operate under the assumption that the same attackers may also bypass their organization's security controls.

\begin{figure}[t]
    \centering
    \includegraphics[width=0.9\columnwidth]{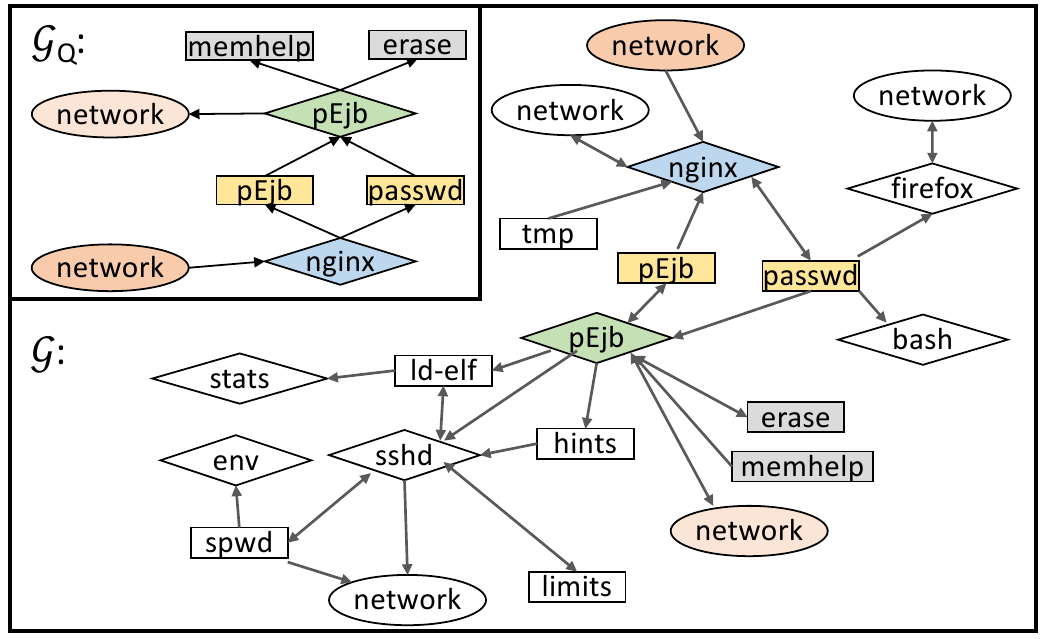}
    \caption{\textcolor{black}{Illustration of the search for threat behaviors in the system-level provenance graph as a subgraph matching problem. 
    $\mathcal{G}_Q$ represents the query graph generated from the described interactions in a threat intelligence text, and $\mathcal{G}$ displays part of the system-level provenance graph centered around the process that executed the malicious code. Matching nodes are colored with the same colors, while unrelated nodes are uncolored or omitted for clarity.}}
    \label{fig:teaser}
\end{figure}

Consider a scenario where a threat hunter discovers news of a new attack targeting an 
\textcolor{black}{organization operating in the same industry as theirs.}
In this case, the appropriate course of action is to hypothesize that the attackers may have already infiltrated their systems and to search for traces of an ongoing intrusion in their system logs.
In the context of provenance analysis, this necessitates converting an externally observed threat behavior into a query that can be searched within system-level provenance graphs \cite{zong2015behavior,milajerdi2019poirot,satvat2021extractor,wei2021deephunter}.  
This problem setting can indeed be viewed as an instance of the graph entailment (subgraph matching) problem. 
\textcolor{black}{To better illustrate our use case, we present an example scenario in Fig. \ref{fig:teaser}.
The scenario involves a threat intelligence report \cite{darpa-2018}, which describes how an adversary compromises a \texttt{nginx} web server by downloading a malicious payload and executing to gain root privileges. 
A threat hunter leverages this information to create a query graph $\mathcal{G}_Q$, where each type of system entity is represented with a distinct shape and color. 
The objective is to search the extensive system-level provenance graph $\mathcal{G}$ and identify nodes that match the query graph in terms of both color and connection pattern.
In this example, the larger subgraph shows where such an alignment is possible, and $\mathcal{G}_Q$ can be confirmed as a subgraph of $\mathcal{G}$, enabling the threat hunter to infer that the described threat behavior is present within that system. 
}

In this work, we propose a solution to this problem and develop a system named \systemNoSpace, which can efficiently and accurately identify matching subgraphs within a large provenance graph corresponding to a given query graph.
The subgraph matching problem involves determining whether a query graph is isomorphic to a subgraph of a larger graph both structurally and in its key features. 
This is a fundamental problem in graph analysis 
with significant practical implications across a broad range of application domains. 
However, exact subgraph matching is known to be NP-complete\footnote{https://en.wikipedia.org/wiki/Subgraph\_isomorphism\_problem}.
Thus, exact methods based on combinatorial approaches are unable to solve large-scale instances.
To perform subgraph matching within a reasonable time frame, several approximate solutions have been proposed. 
Earlier inexact subgraph matching methods typically relied on heuristics to discern better alignment of nodes \cite{liu2019g, liao2009isorankn} or translated the discrete optimization problem into a continuous one \cite{lu2016fast, tong2007fast, tian2007saga}.
Recently, graph neural networks (GNNs) have achieved significant success in graph representation learning~\textcolor{black}{\cite{chen2020graph}}\footnote{\textcolor{black}{In the cybersecurity context, graph representation learning has already demonstrated notable advancements and widespread application, particularly in the domain of vulnerability detection \cite{diwan2021representation, bilot2023survey}}.}.
This led to the development of several learning-based methods for approximate subgraph matching with superior performance \cite{bai2019simgnn,li2019graph,lou2020neural,lan2021sub,roy2022interpretable}. 
At the core of these methods is the learning of an embedding function that maps each graph into an embedding vector encapsulating its key features.
The subgraph relation between two graph embeddings is then evaluated in this continuous vector space.

In the realm of provenance graph analysis, approximate subgraph matching methods encounter distinct challenges 
absent in other domains where they have been successfully implemented \cite{wu2019relation, xu2019cross,davitkova2021lmkg}.
\textcolor{black}{(The characteristics of graph datasets used in this work are presented in Table \ref{tab:graph_size} of Appendix Sec. \ref{sec:graphDatasets}.)}
Provenance graphs are characterized by a large number of nodes and edges, as well as a high average node degree,
due to the diverse activities inherent within a typical computing system.
This results in a considerable computational burden when searching behaviors and learning graph relationships.
Moreover, the coarse-grained nature of logs hinders precise tracking of information and control flows among system entities, leading to erroneous connections between nodes.
These factors render search methods based on node alignment between graphs largely impractical.
 Applying learning-based methods, based on GNNs, to large graphs introduces further complications.
GNNs carry out computation through a series of message-passing iterations, during which nodes gather data from neighboring nodes to update their own information.
The updated information of all nodes is then pooled together to create a graph-level representation.
In this context, increasing the model depth beyond a few layers (i.e., the number of iterations) to more effectively capture relationships results in an exponential expansion of a GNN's receptive field, 
which consequently leads to diminished expressivity due to oversmoothing \cite{loukas2019graph,zeng2021decoupling}.

To improve the efficiency of provenance graph analysis, several methods have been proposed for simplifying provenance graphs, including entity pruning \cite{lee2013loggc,xu2016high, alsaheel2021atlas, hassan2020tactical}, removal of redundant semantics \cite{xu2016high,hossain2018dependence,michael2020forensic,fang2022back,zhu2023aptshield}, behavior abstraction \cite{hassan2018towards,zeng2021watson,xu2022depcomm} and dependence explosion mitigation \cite{lee2013high,ma2017mpi,hossain2020combating,hassan2020omegalog}. 
These data reduction methods primarily focus on identifying anomalous interactions and preserving forensic tractability. 
Nevertheless, they often do not meet the objective of preserving sufficient integrity to support the search for more general graph patterns.

{\bfseries Previous Hypothesis-Driven Threat Hunting Techniques and Limitations:} 
Enhancing threat hunting capabilities necessitates tackling two distinct challenges: effective search and query generation. 
Milajerdi et al. \cite{milajerdi2019poirot} focused on the search aspect by proposing a non-learning-based method, called Poirot,  
to facilitate the search of provenance graphs using known APT attack behaviors as query graphs. 
To reduce search complexity, they adopted a heuristic that assumes an attacker's limited ability to exploit multiple vulnerabilities, allowing them to exclude graph paths without attack nodes.
A significant drawback of this approach is that the entire search computation must be performed at the query time. 
In situations where the nodes of the query graph are commonly present in the target provenance graph, 
the search must consider numerous potential alignments, which may result in covering a large portion of the graph with each query. 

In a similar vein, Wei et al. proposed DeepHunter in \cite{wei2021deephunter}, a learning-based subgraph matching technique that leverages Neural Tensor Networks (NTN) to model the subgraph relationship between two graph-level embeddings and to calculate a matching score. Their method identifies subgraphs surrounding indicators of compromise (IoCs) related to the query, \textcolor{black}{effectively using the query as a filter to reduce the provenance graph}. A match is determined by exhaustively comparing embedding similarities through NTN's pairwise comparisons between the query embedding and sampled subgraph embeddings. 
The efficiency of this method declines due to its dependency on the query, particularly as the size of the provenance graph increases.

The other aspect of hypothesis-driven threat hunting involves creating queries that correspond to specific behaviors to be investigated.
In contrast to the methods used in studies \cite{milajerdi2019poirot,wei2021deephunter}, which relied on manually generated query graphs from incident reports,  Zong et al. \cite{zong2015behavior} tackled this as a query discovery problem.
They developed a discriminative subgraph pattern mining technique to generate query templates automatically.
Alternatively, Satvat et al. \cite{satvat2021extractor} employed a natural language processing pipeline to identify relationships between system entities in threat intelligence reports, ultimately extracting query graphs using this information. 
We focus on the first aspect of the problem, assuming that queries are already available.

{\bfseries Approach Overview and Summary of Contributions:} 
The design of \systemNoSpace\footnote{The source code and trained models associated with our study are accessible online: Anonymized for review.}
addresses the challenge of efficiently processing numerous queries within extensive provenance graph repositories.
\textcolor{black}{As a departure from earlier proposed hypothesis-driven threat hunting methods, our technique leverages a graph representation learning approach that enable the bulk of search computation to be conducted offline, accompanied by a lightweight comparison step performed during query time.}
To effectively learn the diverse range of relations in a provenance graph and mitigate the complexities introduced by its size, we conduct a process-centric partitioning of provenance graphs. Additionally, we implement a behavior-preserving graph reduction method, which incorporates graph versioning to integrate timing information.
The central component of our search methodology is the embedding of subgraphs in a vector space, where subgraph relationships can be directly evaluated.
To achieve this, we employ order embeddings, which allow learning hierarchical entity representations while maintaining the hierarchical structure through the coordinate-wise ordering of these representations in the embedding space. 
Consequently, \system reduces subgraph matching to simple comparisons between query and precomputed subgraph embeddings.
This approach eliminates the necessity for exhaustive computation of all pairwise interactions between query and target subgraph embeddings.
Overall, our study offers significant contributions in the following areas:
\begin{itemize}
\item A graph simplification strategy that preserves the diverse range of behaviors present in a provenance graph, thereby fostering effective learning and search capabilities.
\item The use of order embeddings to facilitate the efficient evaluation of subgraph relationships in the embedding space. 
\item A versatile search methodology that is not exclusively biased towards attack behavior but can be generalized to identify any type of behavior.
\item The ability to efficiently search vast quantities of historical log data, owing to the compactness of subgraph representations and the simplification of provenance graphs. 
\item A substantial decrease in false-matching rates and improved accuracy compared to other hypothesis-driven threat hunting approaches.
\end{itemize}

\section{Challenges in Learning to Search Provenance Graphs}
\label{sec:challenges}
The strength of the provenance graph lies in what it reveals about contextual relationships between system events. 
By mapping the recorded interactions in individual audit log events onto a graph that shows the chronological interplay between processes and other system resources, such as files, network sockets, memory, and registry objects, in the form of system calls, they allow for the use of graph analysis techniques. 
In this regard, a provenance graph is a heterogeneous, typed, directional, and dynamic graph that provides a coarse-grained insight into  system state changes.
This rich representation, however, also poses a number of key challenges for graph-learning methods.

{\bf \textcolor{black}{Challenge \#1:} Size of Graphs.}
Many system events are recurrent, in mundane nature, and affect a large number of system objects such as system and software updates, backup, and data synchronization jobs. \eat{[Other EXAMPLEs]}
As a result, audit logs collected from a typical machine may easily yield provenance graphs with a very large  number of nodes and edges over shorter durations of time~\cite{xie2018pagoda}.
The expressive power of a GNN is determined by its capacity, which is generally expressed in terms of the width and depth of a neural network, i.e., the embedding size and the number of layers of a GNN, respectively. 
Loukas \cite{loukas2019graph} studied the difficulty of well-known subgraph analysis tasks and determined that even  verifying whether a subgraph meets a given property requires the product of a GNN’s depth and width higher than a (low-order) polynomial of the size of a graph. 
This lower bound implies that for large query and provenance graphs learning subgraph relations will indeed be difficult.
Further, for large graphs increasing the model depth, i.e., the number of layers in a GNN, 
often translates to exponential expansion in the number of neighboring nodes.
This implies that the scope (i.e., receptive field) of a GNN has to be restricted through methods such as subgraph-based sampling \cite{chiang2019cluster, zeng2019graphsaint} when learning subgraph relationships.

{\bf \textcolor{black}{Challenge \#2:} High-Degree Nodes.}
The average degree of a node in the provenance graph can be quite high.
For example, the IP node of a DNS server may have incoming and outgoing node degrees that easily exceed tens of thousands, while a /dev/null device might have incoming node degrees reaching several thousands.
This is partly due to the dependence explosion problem, where long-running processes interacting with many other system entities appear as highly connected nodes on the graph. 
The computation in GNNs is performed through several message-passing iterations in which nodes aggregate information from adjacent neighbors to update their information. 
For high-degree nodes, this aggregation step is likely to suppress useful characteristics. It will result in degraded expressivity due to the  well-known over-smoothing behavior where node embeddings become uninformative after several rounds of message-passing \cite{huang2020tackling}. 
This is especially a concern for system behaviors that occur infrequently. 
This indeed requires adopting a representation that exhibits system behaviors in a balanced manner.

{\bf \textcolor{black}{Challenge \#3:} Preserving Time Order of Events.}
Edges in a provenance graph represent a time-ordered sequence of events. 
Disregarding the timing information essentially introduces spurious information flows among system entities.  
The computation in GNNs includes the creation of computation graphs rooted at each node describing the structure for message-passing and aggregation.
The timing information can be incorporated during graph creation by enforcing a time order going from root to leaves.
In practice, however, a recursive neighborhood aggregation scheme is used to avoid the computational overhead of repeated creation of these computation graphs. 
This scheme cannot preserve the causality relation between edges as all nodes concurrently aggregate information from their neighbors.
In addition, it needs to be considered that a queried behavior may not explicitly include the timing information between depicted events. 
Therefore, timing information must be utilized so that subgraph relationships do not explicitly depend on it. 

{\bf \textcolor{black}{Challenge \#4:} Semantic Gap with Query.}
Another challenge is the potential mismatch in the degree of expressiveness between the queried behavior and its observed version on the provenance graph.
An essential source for threat behavior includes incident response reports where an analyst gathers evidence related to steps of malicious activity on a system \cite{satvat2021extractor}. 
In this regard, the query graph may not necessarily convey the system-level interactions with all its details.
For example, a query graph may show a browser or application process as part of an attack vector. 
However, on the audit logs, this process may correspond to a cloned version of the main process or another process spawned by a launcher process to handle the task, which 
may not be a part of the query.
Additionally, missing or unrecorded events or the inclusion of additional interactions can cause the queried behavior to only partially match its version on the provenance graph.
Thus, the learned representation must offer a degree of robustness to such variations.

{\bf \textcolor{black}{Challenge \#5:} Setting a Learning Objective.}
The learning objective serves as a guiding principle for the model's training process. 
Establishing a clear and well-defined learning objective is essential for the model to effectively learn subgraph relations in query graphs.
This entails generating training samples that accurately represent the characteristics of the query graphs that are expected to be encountered in practical usage. 
To promote model generalization, it is essential to draw training and test samples from the same data distribution. This consistency allows the model to effectively learn the underlying patterns and relations in the data, enhancing its ability to recognize subgraph relations in new and unseen data.
In the context of threat hunting, query graphs are anticipated to exhibit a strong relationship with observed attack behaviors. Consequently, it is vital to adapt the model's training process to reflect these behaviors.

\section{System Design and Methodology}

Our approach to hypothesis-driven threat hunting utilizes provenance graphs and frames the task as a subgraph entailment problem. 
Provenance graphs depict audit logs as labeled, typed, and directed graphs, where nodes represent system entities and directed edges indicate transformation or flow of information due to distinct system calls.
Timestamps assigned to each node and edge capture the graph's evolving nature.
Our technique aims to effectively identify system behaviors of interest by representing queries as graphs and searching for them within the larger context of the provenance graph.

\begin{figure*}[!]
    \centering
    \includegraphics[width=0.75\textwidth]{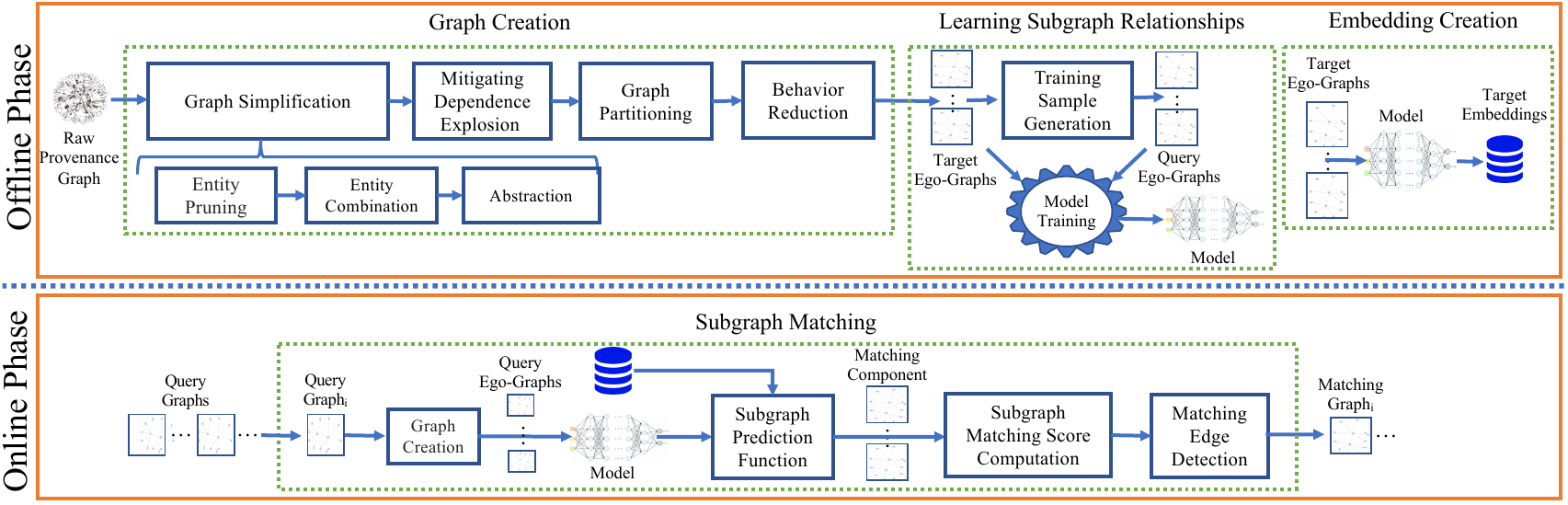}
    \caption{Overview of \system and its key components.}
    \label{fig:overview}
\end{figure*}

\subsection{Problem Formulation}
A graph $\mathcal{G}$ is defined as a set of nodes $\mathcal{V}=\{v_1,...,v_N\}$ and a set of edges $\mathcal{E}=\{e_1,...,e_M\}$ where each node and each edge are associated with a type. 
Given a target and a query graph, the solution to the graph entailment problem involves detecting every query instance in the target.
Since exact subgraph matching on graphs with the scale of provenance graphs is not feasible,
we employ an approximate matching method to make subgraph predictions. 
Our method involves a sequence of steps that reduce the size of a graph while ensuring that the system behavior is preserved at a higher level of abstraction.

Following this reduction, the primary problem we want to solve is formulated as follows.
Let $\mathcal{G}$ represent a reduced provenance graph.
We decompose $\mathcal{G}$ into a set of overlapping subgraphs by extracting the $k$-hop ego-graph\footnote{\textcolor{black}{An ego-graph of depth $k$, centered around a node $p$, is an induced subgraph that includes $p$ and all nodes within a distance $k$ from it.}} $\mathcal{G}_{p}$ of each process node $p\in \mathcal{V}_P$.
Given the set of ego-graphs $\mathcal{P}=\{\mathcal{G}_{p} | \mathcal{G}_{p} \subseteq \mathcal{G}\}$, our goal is to learn an embedding function $\eta : \mathcal{G}_{p} \rightarrow  \mathbb{R}^d$ that maps each ego-graph to a d-dimensional representation $\mathbf{z} \in \mathbb{R}^d$, capturing the key structural information of a graph for use in conjunction with a suitable subgraph prediction function $\varphi$.
Hence, the encoder must incorporate an inductive bias to effectively represent the subgraph relationship while learning a mapping
in which the subgraph prediction function $\varphi(\mathbf{z}_p,\mathbf{z}_q)$ serves as a vector-based measure to confirm the existence or absence of this relation.
It must further be noted that since provenance graphs are typically very large, $\varphi(\mathbf{z}_p,\mathbf{z}_q)$ needs to be evaluated over all $\mathbf{z}_p$ values for a given $\mathbf{z}_q$.
Therefore, effective computation of $\varphi(\mathbf{z}_p,\mathbf{z}_q)$ is very critical.

In our method, we employ a subgraph embedding function that effectively addresses both issues. 
For this, we utilize the notion of order embedding, which aims to encode the ordering properties among entities into target representation space \cite{vendrov2015order,athiwaratkun2018hierarchical,vilnis2018probabilistic,chiang2019one}.
Order embeddings specifically model hierarchical relations with anti-symmetry and partial-order transitivity, which are inherent to subgraph relationships. 
To develop the embedding function $\eta$, we utilize an inductive graph neural network \cite{graphsage:nips:2017} and apply the order embedding technique introduced by Vendrov et al. \cite{vendrov2015order}.
This approach enables us to learn a geometrically structured embedding space that effectively represents the relationships between subgraphs.
At the query execution state, the encoder $\eta$ is applied independently to the query graph $\mathcal{G}_{Q}$. This is done by identifying ego-graphs
$\mathcal{Q}=\{\mathcal{G}_{q} | \mathcal{G}_{q} \subseteq \mathcal{G}_{Q}\}$ corresponding to all anchor nodes $q\in \mathcal{V}_{Q}$ and computing the embeddings $\mathbf{z}_q= \eta(\mathcal{G}_{q})$ for all ego-graphs in $\mathcal{Q}$.

Then, the subgraph prediction function is evaluated by considering the newly computed embeddings $\mathbf{z}_q$ from the query and the precomputed subgraph embeddings $\mathbf{z}_p$. This involves identifying $(p, q)$ node pairs that satisfy the subgraph relation $\varphi(\mathbf{z}_p,\mathbf{z}_q)$.
To determine whether one graph is a subgraph of another, one can simply check that all neighbors of $q \in V_Q$ satisy the subgraph relationship. 
However, such a comparison enforces an exact match of the query, which cannot handle cases where discrepancies exist between the query and the logs. 
To address this issue and achieve greater generality, we propose the use of a soft-decision metric, defined as follows:
\begin{eqnarray}
&&\mathcal{G}_{Q} \subseteq \mathcal{G}_{P}\, \text{ iff }\, g(\mathcal{G}_{p}^{*}, \mathcal{G}_Q) \geq \tau \text{ where } \nonumber  \\ 
&&\mathcal{G}_{p}^{*}=\{\cup\mathcal{G}_p | \exists q \in V_Q, \,  \varphi(z_p, z_q)\ \text{is satisfied}\}
\label{eq:match}
\end{eqnarray}
Here, $\mathcal{G}_{p}^{*}$ represents a graph obtained by combining all ego-graphs $\mathcal{G}_p$ that satisfy the
subgraph relationship with the query ego-graphs of $\mathcal{G}_Q$, and $g(.,.)$ is a scoring function that computes the intersection of 
 $\mathcal{G}$ and the query graph $\mathcal{G}_Q$.

\subsection{System Overview}
Our system, displayed in Fig. \ref{fig:overview}, consists of an offline embedding stage and an online prediction stage. 
During the offline phase, audit logs are collected and used to create a provenance graph. 
Several graph simplification steps are taken to optimize graph learning and better align with queried behavior (Sec. \ref{subsec:graph_simplification}). 
These involve removing non-essential nodes and edges related to pipe and memory accesses, merging cloned processes with their parents, and consolidating  network communication events over time intervals. 
Each node that describes a unique system entity is then assigned to an abstract class that describes a higher-level categorization, such as a
system file, a user application process, etc.
After simplifying the graph, the next step is to version the nodes, which incorporates timing information into the graph to prevent spurious information flows between nodes (Sec. \ref{subsec:dependence_explosion}).
In the final step, the graph is partitioned into overlapping subgraphs by extracting the $k$-hop ego-graph of each process node.
Here, $k$ also signifies the number of GNN layers used to obtain a subgraph representation.
We then apply another level of reduction to the abstracted ego-graphs by removing repeated behaviors using an iterative label propagation  (Sec. \ref{subsec:behavior_reduction}).
This leaves unique traits in each subgraph to learn as part of a subgraph relation.

We then employ a $k$-layer GNN to learn a representation of the subgraph relation by training it on positive and negative pairs of graphs to learn an inductive embedding function that will be used in conjunction with a subgraph prediction function.
For this, we utilize order embeddings which provide a natural way to model transitive relational data such as entailing graphs.
These embeddings essentially obey a structural constraint whereby $\mathcal{G}_q$ is deemed a subgraph of $\mathcal{G}_p$ if and only if
all the coordinate values of $\mathbf{z}_p$ are higher than $\mathbf{z}_q$'s (Sec \ref{subsec:subgraph}).
During the prediction stage, the query graph undergoes the same processing steps as the provenance graph and is partitioned into subgraphs.
Afterward, the order relations between the query ego-graph embeddings and the precomputed ego-graph embeddings in the provenance graph 
are computed to determine whether the subgraph relation exists (Sec. \ref{subsec:matching}).

\subsection{Threat Model}
\textcolor{black}{
We assume that adversaries cannot tamper with the system or the kernel auditing facility responsible for collecting security logs, ensuring the integrity of the constructed provenance graphs.
However, adversaries who possess knowledge of our system might make slight adjustments to their behavior in an attempt to better conceal their traces.
To obscure their actual behavior from threat hunters, an adversary may choose to refrain from performing specific steps or replace some of the attack steps with alternative ones. 
Hence, it is crucial for our system to exhibit robustness against the insertion and deletion of nodes and edges in the query graph representing the attack behavior. 
Although limiting the impact of such modifications poses a significant challenge, it is widely recognized that adversaries face difficulties in frequently altering their tactics, techniques, and procedures. 
Moreover, engaging in arbitrary activities during an attack would increase the risk of detection by other system-wide attack detection tools.
As a result, we proceed with the assumption that the fundamental characteristics of the attack behavior will remain unchanged and that similarity will be largely preserved.
}

\section{System Components} 

\subsection{Graph Creation }
\label{sec:graph_creation}
\system processes raw system logs through multiple reduction steps before constructing a streamlined provenance graph that represents various interactions
between subjects (e.g., processes) and objects (e.g., processes, files, network sockets). 
The graph creation module produces a partitioned version of this graph to facilitate effective learning of subgraph relationships.

\subsubsection{Graph Simplification (\textcolor{black}{Challenges \#1 and \#4})}
\label{subsec:graph_simplification}

We start the process by adopting the approach from previous research ~\cite{provdetector2020, wei2021deephunter}, where we retain only process, file, network, and registry objects. 
We maintain all read, write, and modify attribute events for processes, files, sockets, and registries. Additionally, we preserve clone, fork, or execute events for processes, while removing open and close events to avoid redundancy, as they precede or follow read or write events.

\begin{figure}[!ht]
\centering
    \includegraphics[width=0.8\columnwidth]{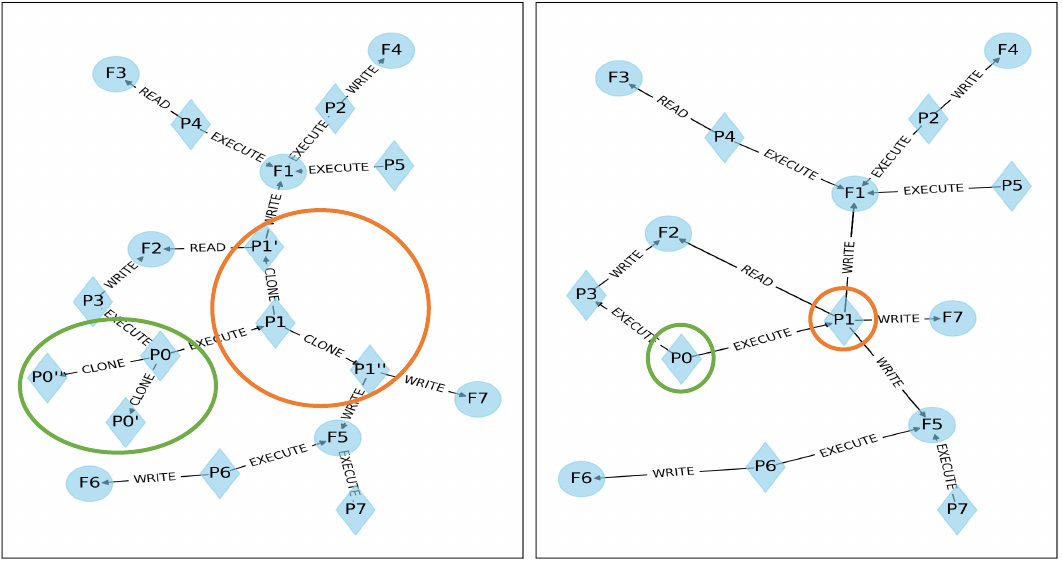}
   \caption{An example provenance graph consisting of processes ($\diamondsuit$) and files ($\circ$). 
   The node labels represent the names of the files and processes. 
   The nodes that represent threads of processes $P_0$ and $P_1$ are surrounded by circles.
    The simplified graph after combining threads is shown on the right side.}
  \label{fig:initial}
\end{figure}

\textcolor{black}{
We expand upon the generic graph simplification approaches in two main aspects. The first aspect pertains to the handling of threads during the creation of provenance graphs.}
Applications often use threads to enhance performance and scalability, but query graphs might not exhibit this behavior~\cite{thalheim2016inspector}. To ensure consistency across both graphs, we merge threads into their parent process
as illustrated in Fig. \ref{fig:initial}. 
Furthermore, to capture changes in the behavior of remote servers over time, we adopt the method described in \cite{hossain2018dependence}, 
treating each remote IP and port combination as a distinct source within 10-minute time windows.

\textcolor{black}{The second aspect focuses on achieving a consistent representation of system entities and their interactions in both the query and provenance graphs.}
A discrepancy in these representations could hinder the model's generalization capability.
This is particularly concerning when the query lacks the granularity and detail typically found in system logs, potentially leading to mismatches or misinterpretations when comparing the query and provenance graphs.
For example, a browser process in the query may correspond to one of several processes, such as Firefox, Chrome, or Safari, in the provenance graph. 
Similarly, two files with the same name may be associated with different functions in the context of different applications.
To avoid such ambiguities, we use a system directory-based abstraction for all system entities, with the exception of network objects, in order to provide a consistent description for each entity.
Specifically, we assign category labels to each entity based on its root directory in the file system, indicating a higher-level function for each entity.
In contrast, network objects are abstracted based on their source IP, destination IP, and destination port. Each IP address is categorized as \textit{public, private}, or \textit{local} based on its usage, while ports are categorized as \textit{reserved} if they are less than 1024 and as \textit{user} otherwise. 
Overall, this led to the use of more than 70 abstraction categories, which are summarized in Appendix Sec. \ref{sec:abstract_category}.

Abstracting system entities not only helps reveal recurring patterns in a graph but also allows for further reduction in graph complexity. When object nodes—such as files, network sockets, and registry entries—within the same abstraction category are connected to a single subject (process) node through a shared event type, we merge these nodes into one node with the same object abstraction. To maintain causality relationships during deduplication, we keep the timestamp of the first event if the flow starts from a process to an object (e.g., write or attribute modification) and the timestamp of the last event if the flow originates from an object and leads to a process. Since these nodes are connected to only one process, this procedure preserves the causality relationships between nodes.

\subsubsection{Mitigating Dependence Explosion (\textcolor{black}{Challenges \#2 and \#3)}}
\label{subsec:dependence_explosion}
The dependence explosion, caused by high in-degree and out-degree nodes in a provenance graph, 
significantly impedes the learning of subgraph relationships.
This is because tracing through such nodes leads to an exponential increase in the possible node interactions that must be considered.
We implement two strategies to address this problem while extracting ego-graphs from the provenance graph.
First, we leverage the available event timestamp information to impose a timing constraint on the flows.
To satisfy this requirement, we employ the graph versioning approach proposed in \cite{hossain2018dependence}, which effectively encodes time dependencies into the provenance graph by creating a new version of a node whenever the corresponding system entity receives new information. 
This method ensures that all paths in the extracted ego-graphs have edges with monotonically increasing timestamp values, thereby preserving the causal order of events.
Additionally, it allows the elimination of repeated events between two versioned nodes.
It's worth noting that incorporating node versioning in provenance graphs does not necessitate the inclusion of edge timestamps in the query graph.

\textcolor{black}{The second strategy we propose aims to address the challenges of oversmoothing in GNNs by alleviating the impact of dependence explosion.}
For this, we designate specific nodes as sink nodes. 
Notably, interactions with high-degree nodes, such as DNS server IP addresses or cache files,  do not provide discriminative information that aids in learning subgraph relationships.
Moreover, any system entity interacting with these high-degree nodes will appear to receive information from numerous other system entities. 
This contributes to the oversmoothing phenomenon, as it results in an expanded receptive field for a GNN.
By treating these nodes as sink nodes, we effectively prevent non-informative information flows, leading to more accurate and meaningful learning of subgraph relationships.
It is important to note that non-process nodes with zero in-degree or out-degree, such as log files written to by all processes without reads, or configuration files that are only read, are also considered sink nodes.
This is because there is no flow of information between the neighbors of these nodes, making their role in understanding subgraph relationships less significant.

\subsubsection{Graph Partitioning (\textcolor{black}{Challenge \#1)}} 
\label{subsec:graph_partitioning}

A pattern within a graph can always be detected in a sufficiently large local neighborhood surrounding a specific node. 
An ego-graph with depth $k$, centered around node $v$, is an induced subgraph that includes $v$ and all nodes within a distance $k$ from it. 
\textcolor{black}{In fact, any pattern with a radius smaller than $k$ can be found within an ego-graph of depth $k$, where the value $k$ can be determined based on the characteristics of query graphs and the graph representation learning algorithm.} 
To take advantage of this, the reduced provenance graph, containing versioned and sink nodes, is partitioned into subgraphs by extracting the ego-graph of each process node which are crucial for understanding the system behavior~\cite{thalheim2016inspector}.
\textcolor{black}{
Given that processes are the only active system entities, adopting a process-centric view of the provenance graph indirectly encompasses all relationships involving other entities. This approach also contributes to reducing the computational complexity of both offline and online stages.}
Even though the graph partitioning step is performed once, 
it is essential for this task to be as efficient as possible.
To extract ego-graphs, we use a dynamic programming algorithm presented in the Algorithm \ref{alg:partitioning} in Appendix Sec. \ref{sec:A1}.

\begin{figure}[!t]
\centering
    \includegraphics[width=0.8\columnwidth]{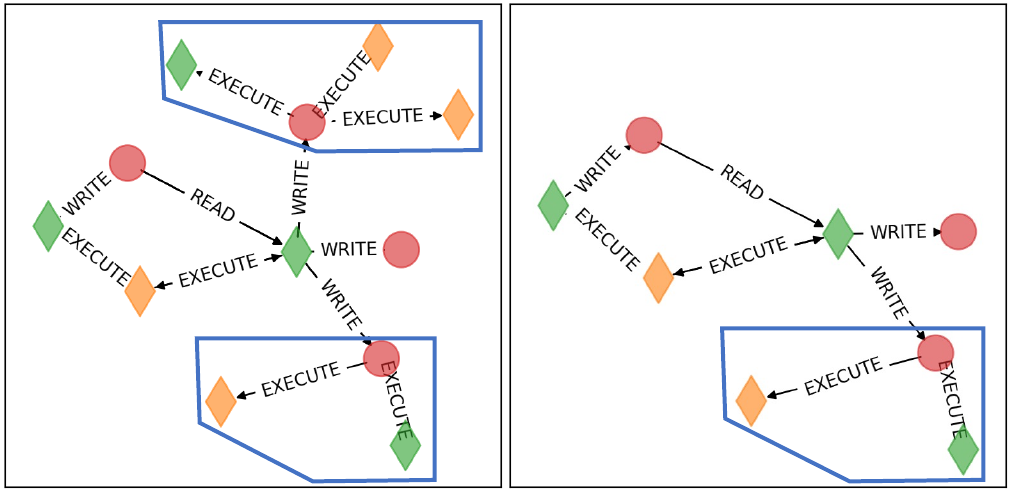}
   \caption{
   The left figure displays the ego-graph of a process $P_1$ after applying simplification and dependence explosion steps to the initial graph shown in Fig. \ref{fig:initial}. The right figure presents the resulting graph after applying behavior-preserving reduction. Algorithm \ref{alg:reduction} is utilized to identify the recurring behavior, as indicated within the blue boxes.}
  \label{fig:reduction}
\end{figure}

\subsubsection{Behavior-Preserving Graph Reduction (\textcolor{black}{Challenges \#1 and \#4)}} 
\label{subsec:behavior_reduction}

At this stage, resulting ego-graphs may still contain redundant information. 
For instance, consider an ego-graph showing a user process that has written to hundreds of \texttt{var} directory files, possibly in different contexts. 
A given query involving this user process, however, is likely to be relevant to only one of these contexts.
Therefore, from a search perspective, a user process writing to a \texttt{var} directory is more informative than tracking the number of written files. 
Moreover, as repeated events can dominate the information aggregation step, GNNs may primarily learn those repetitive behaviors while neglecting less frequent ones.
To avoid the suppression of observed system behaviors, it is necessary to identify and eliminate recurring patterns within each ego-graph\footnote{\textcolor{black}{A similar approach was taken by Watson \cite{zeng2021watson} to cluster similar system behaviors on the provenance graph. However, unlike \systemNoSpace, Watson adopts a sequence modeling approach by decomposing subgraphs into paths and utilizing TransE knowledge embeddings.}}.

We propose a behavior-preserving graph reduction method that captures the local graph structure surrounding each node at varying depths using iterative label propagation.
First, we collapse the node versions back onto the original nodes.
(While this loses the benefits of versioning, maintaining time-dependency during ego-graph creation ensured that only relevant system entities are included in each ego-graph, eliminating all spurious interactions with unrelated system entities.)
We compute a hash value for each node by aggregating edges with their neighboring nodes' hash values.
For each node, the abstraction category of the node is assigned as its $0$-hop hash, and the hashes of neighboring nodes are accumulated from incoming and outgoing edges using the following equation:
\begin{eqnarray}
\label{eq:hash}
nh[n][forw][l] = hash \left (\sum_{e, v \in In(n)} \left (e + nh[v][forw][l-1] \right) \right),\nonumber\\
\label{eq:hash}
nh[n][back][l] = hash \left (\sum_{e, v \in Out(n)} \left (e + nh[v][back][l-1] \right) \right).\nonumber
\end{eqnarray}
Here, $nh[n][forw][l]$ and $nh[n][back][l]$ represent the hash values of node $n$ at $l$-hop distance in the forward and backward directions, respectively. 
The function $hash()$ is the SHA-256 function that takes a string as an input and returns a cryptographic hash value. 
We use the function $In(n)_e$ to retrieve all incoming edges with their source nodes for a given input node $n$, while $Out(n)_e$ returns all outgoing edges and their target nodes. 
The set function $\sum$ returns the unique strings in its input in sorted order.

We perform behavior-preserving reduction starting from the anchor node of an ego-graph. 
At each depth $l$, we determine all unique $(k-l)$ hash values of the neighbor nodes and select one node for each unique hash value. 
These selected nodes form the set of unique nodes for that depth. 
We repeat this process at all depths up to $k$ and obtain a set of unique nodes for the entire ego-graph. 
Using these unique nodes, we create a reduced ego-graph that preserves the behavior of the original one. 
The detailed steps of our behavior-preserving reduction are provided in Algorithm~\ref{alg:reduction}. The $random\_select$ function takes a list of nodes as input and returns one random node from the input list, and the $subgraph$ function is used to create a reduced graph with the input nodes.
\textcolor{black}{Figure \ref{fig:reduction} depicts the resulting ego-graph obtained after applying the behavior-preserving reduction step.}

\begin{algorithm}[!tb]
\caption{Behavior-Preserving Graph Reduction Method}
\label{alg:reduction}
\begin{algorithmic}[1]
\REQUIRE $\mathcal{G}_{p}$: ego-graph of anchor node $v_p$, $nh$: hashes of node $n$
\ENSURE $\mathcal{G}_{p}$: reduced ego-graph of anchor node $v_p$

\STATE $forward \leftarrow V_p$
\FOR{$l=0,..,k$}
    \STATE $unique \leftarrow dict()$
    \FORALL{$e,v \in In_e(forward)$}
        \STATE $unique[e + nh[v]['forw'][k-l]].append(v)$
    \ENDFOR
    \FORALL{$hash \in unique$}
        \STATE $forward.append(random\_select(unique[hash])$
    \ENDFOR
\ENDFOR
\STATE $backward \leftarrow V_p$
\FOR{$l=0,..,k$}
    \STATE $unique \leftarrow dict()$
    \FORALL{$e,v \in Out_e(backward)$}
        \STATE $unique[e + nh[v]['back'][k-l]].append(v)$
    \ENDFOR
    \FORALL{$hash \in unique$}
        \STATE $backward.append(random\_select(unique[hash])$
    \ENDFOR
\ENDFOR
\STATE $unique\_nodes \leftarrow forward|backward$
\STATE $\mathcal{G}_{p} \leftarrow \mathcal{G}_{p}.subgraph(unique\_nodes)$

\end{algorithmic}
\end{algorithm}

\subsection{Learning Subgraph Relationships}
\label{subsec:subgraph}
\subsubsection{Subgraph Prediction Function (\textcolor{black}{Challenge \#5)}:}
Subgraph relationship essentially imposes a hierarchy over graphs. 
Therefore, a vector representation for subgraphs should take into account the structure of this hierarchy to effectively evaluate the relationship between two graphs. 
In \cite{vendrov2015order}, Vendrov et al. introduced order embeddings to model the transitivity and antisymmetry of partially ordered data,
which naturally applies to the representation of entailing graphs\footnote{Subgraph relation is a partial order relation as it has the following three properties Given three graphs $\mathcal{G}_a$, $\mathcal{G}_b$, $\mathcal{G}_c$, 
subgraph relationship satisfies the following three properties: (i) $\mathcal{G}_a \subseteq \mathcal{G}_a$ (reflexivity) (ii) if $\mathcal{G}_a \subseteq \mathcal{G}_b$ and  $\mathcal{G}_b \subseteq \mathcal{G}_c$, then $\mathcal{G}_a \subseteq \mathcal{G}_c$ (transitivity); and (iii) if $\mathcal{G}_a \subseteq \mathcal{G}_b$ and  $\mathcal{G}_b \subseteq \mathcal{G}_a$, then $\mathcal{G}_a = \mathcal{G}_b$ (antisymmetry).}.
Order embeddings ensure the preservation of partial ordering between elements by maintaining the order relations of coordinates in the embedded space
such that for two graphs $\mathcal{G}_p$ and $\mathcal{G}_q$ and their embeddings $\mathbf{z}_{p}, \mathbf{z}_q \in \mathbb{R}^d$
\begin{equation}
\mathcal{G}_q \subseteq \mathcal{G}_p \text{ if and only if } \forall_{i=1}^d \, \mathbf{z}_{p_i} \geq \mathbf{z}_{q_i}.
\label{eq:order}
\end{equation}
That is, $\mathcal{G}_q$ is a subgraph of $\mathcal{G}_p$ if and only if all the coordinate values of $\mathbf{z}_{p}$ are higher than $\mathbf{z}_{q}$’s.
To impose this constraint on the learned relation, \cite{vendrov2015order} proposed an order violation penalty function 
\begin{equation}
    E(\mathbf{z}_q,\mathbf{z}_p)= \|\max\{0,\mathbf{z}_q -  \mathbf{z}_p\}\|^2
\end{equation}
to measure the extent to which two embeddings violate their order, i.e., $E(\mathbf{z}_q,\mathbf{z}_p)\not=0$ if  Eq. (\ref{eq:order}) is not satisfied.
Consequently, we optimize our GNN to minimize the order violation penalty to learn an approximate order embedding function 
using the following max-margin loss
\begin{equation}
   \mathcal{L}(\mathbf{z}_q,\mathbf{z}_p) = \sum_{(\mathbf{z}_q,\mathbf{z}_p)\in S^+} E(\mathbf{z}_q,\mathbf{z}_p) + 
    \sum_{(\mathbf{z}_{q'},\mathbf{z}_{p'})\in S^-} \max\{0, \alpha-E(\mathbf{z}_{q'},\mathbf{z}_{p'})\} 
    \label{eq:loss}
\end{equation}
where $S^+$ denotes a set of positive graph pairs that satisfy the subgraph relation, and $S^-$ is the set of negative pairs for which this relation is not satisfied.
This loss crucially encourages positive samples to have zero penalty and negative samples to have a penalty greater than a margin $\alpha$,
thereby ensuring that two embeddings have a violation penalty of $\alpha$ if the subgraph relation does not hold.
Thus, the subgraph prediction function introduced in Eq. (\ref{eq:match}) becomes a proxy for the order violation penalty, 
i.e., $\varphi(z_p, z_q)=E(\mathbf{z}_q,\mathbf{z}_p)$.
In our evaluation, as an alternative, we also utilized a neural network model to learn the intrinsic relationship between embeddings $\mathbf{z}_q$ and $\mathbf{z}_p$ of entailing graphs as a representation for $\varphi(\mathbf{z}_p,\mathbf{z}_q)$.

\subsubsection{Training Sample Generation (\textcolor{black}{Challenge \#5)}}
\label{subsec:sample_generation}

Training \system requires positive and negative pairs of query and target graphs.
These pairs can be represented as  $(\mathcal{G}_q^+, \mathcal{G}_p)$ and $(\mathcal{G}_q^-,\mathcal{G}_p,)$ where $\mathcal{G}_q^+$ is a subgraph of $\mathcal{G}_p$ and $\mathcal{G}_q^-$ is not. 
During training, the model first computes embeddings for all graphs in a batch of positive and negative pairs, then evaluates the resulting loss as defined in Eq. (4). We backpropagate this loss to update the network weights and minimize its value.

To ensure generalization, $\mathcal{G}_p$  is chosen as an ego-graph of a node $v_p$ within a reduced provenance graph $\mathcal{G}_\textbf{P}$. 
There are two crucial factors to consider when creating a paired query graph. The first is the size of the query graphs.
In line with previous research
\cite{zong2015behavior,milajerdi2019poirot,wei2021deephunter,satvat2021extractor}, 
we opted to limit the size of reduced query graphs to 10-15 edges considering 3-hop ego-graphs. 
We note that in unreduced query graphs, this may correspond to 40-50 edges as discussed in the findings of Table \ref{tab:effect} in Sec. \ref{sec:res-reduction}.
The second factor is the strategy employed to generate $\mathcal{G}_q^+$ and $\mathcal{G}_q^-$.
The most straightforward approach to create $\mathcal{G}_q^+$ involves subsampling a set of nodes or edges from $\mathcal{G}_p$ and extracting the corresponding node or edge-induced graph. 
However, a random selection scheme could expose the model to repetitive behaviors and lead to overfitting common graph patterns.
As for $\mathcal{G}_q^-$, choosing a graph at random may not only generate easy negative samples but also inadvertently yield an actual subgraph of $\mathcal{G}_p$, particularly when $\mathcal{G}_p$ is large.

To circumvent these pitfalls, we propose a new graph sampling method based on path frequency. 
First, possible flows for each ego-graph, $\mathcal{G}_p\in \mathcal{G}_\textbf{P}$, are determined via forward and backward depth-first search around the anchor node $v_p$, where a flow represents a path between two nodes of $\mathcal{G}_p$ that passes through $v_p$.
To generate positive graph pairs, we count the unique flows for each ego-graph $\mathcal{G}_p$ belonging to the same process path.
Then, for each $\mathcal{G}_p$, we randomly select a flow from all its flows based on their inversely weighted frequency in all ego-graphs of the same path.
Once the flow is selected, we expand it by randomly choosing some incoming and outgoing edges of the nodes in the selected flow until the desired number of edges is reached.

Creating a negative example is more challenging as one needs to avoid introducing both superficial and unlikely behaviors to $\mathcal{G}_q^-$.
One can indeed create a hard negative example $\mathcal{G}_q$ by synthetically adding edges and nodes to a target graph $\mathcal{G}_p$ to violate the subgraph relationship.
However, this may result in implausible behaviors. 
Alternatively, one can identify an arbitrary flow from the list of known unique flows, which is not contained within the target $\mathcal{G}_p$, and use the corresponding process's ego-graph to expand this flow which may result in a very easy example for the model. 
Instead, we follow a three-step procedure to create a negative sample: 
First, we pick a flow from an ego-graph with the same anchor process as the target graph and expand from it. 
(E.g., to create a negative example for a Firefox process, we prefer to choose an ego-graph of another Firefox process and subsample it.)
However, this may not always be possible if there are not many instances of the same process.
Because in such a case, the same behavior may potentially be used to generate many negative examples, thereby biasing the model.
To avoid this, as a second step, we utilize the behavior of another process with the same abstraction, i.e., using a Chrome process instead of a Firefox.  
Where this is not possible, as a last resort, we pick a random flow and expand from it.

To ensure that the generated ($\mathcal{G}_q^-$, $\mathcal{G}_p$) pairs violate the subgraph relationship, we apply an independent validation step. 
For this, we first check if any node or edge abstraction is present in $\mathcal{G}_q$ but absent in $\mathcal{G}_p$.
If all categories of system entities are indeed found within $\mathcal{G}_p$, we proceed to analyze all 1- and 2-hop flows in the query graphs, taking both edge types and node abstractions into account. 
Should at least one distinct flow fail to meet the subgraph relationship criteria, the pair is deemed a negative sample.
Sample positive and negative queries are presented in Appendix Sec. \ref{sec:app-query}.

\subsubsection{GNN Architecture and Features}
Graph neural networks (GNNs) are expressed as message-passing networks that rely on three key functions, namely \texttt{MSG}, \texttt{AGG}, and \texttt{UPDATE}. These functions work together to transfer information between the different components of the network and update their embeddings. 
Typically, these functions operate at the node level, exchanging messages between a node $v_i$ and its immediate neighboring nodes $\mathcal{N}_{v_i}$. 
In layer $l$, a message between two nodes $(v_i, v_j)$ depends on the previous layer's hidden representations $h_i^{l-1}$ and $h_j^{l-1}$, i.e, $m_{ij}^l = \texttt{MSG}(h_i^{l-1}, h_j^{l-1})$. 
Then, \texttt{AGG} combines the messages from $\mathcal{N}_{v_i}$ with $h_i^{l-1}$ to produce $v_i$'s representation for layer $l$ in \texttt{UPDATE}. 
Various adaptations of this core message passing framework with alternative \texttt{MSG}, \texttt{AGG}, and \texttt{UPDATE} implementations have been proposed~\cite{gcn:iclr:2017, graphsage:nips:2017, gat:velickovic2018graph, hgt:web:2020}. 
To leverage the comprehensive representation of provenance graphs, we deploy a multi-relational GNN that can also incorporate information  by taking into account both edge type and edge direction relations
~\cite{rgcn2018}.

The expressive power of GNNs is known to increase when node and edge features become more distinct~\cite{hamilton2017representation}. 
To take advantage of this, it is essential to assign suitable node and edge features. 
We employ two separate one-hot encoding representations for each object type and abstraction category, and the node features for both the provenance and query graphs are determined in the same way.

\subsection{Subgraph Matching Score Computation}
\label{subsec:matching}

Our technique relies on two measures to achieve robustness against inexact queries, in cases where the query may not precisely match the system events being searched for.
The first measure is utilized when assessing the subgraph relationship between two ego-graphs, as defined in Eq. (5), by permitting a certain degree of order violation, i.e., $\varphi(z_p, z_q)=E(\mathbf{z}_q,\mathbf{z}_p)\leq \tau_{ovp}$. 
The second measure allows for partial matching of the query graph within the provenance graph, which is achieved by using a graph intersection-based scoring function.
The graph $\mathcal{G}^*$, as described in Eq. (\ref{eq:match}), is the union of all possible matches $\mathcal{G}_p$ to $\mathcal{G}_Q$ and may contain several disconnected parts. 
The scoring function intersects the query graph with each connected component (CC) of $\mathcal{G}^*$ and utilizes the ratio of edges in the intersected graphs to the total number of edges in $\mathcal{G}_Q|$ to compute the final matching score, as defined below:
\begin{eqnarray}
g(\mathcal{G^*}, \mathcal{G_Q})=max \left(\left\{\frac{|\mathcal{G}_1^*\cap \mathcal{G}_Q|}{|\mathcal{G}_Q|}, \ldots, 
\frac{|\mathcal{G}_n^* \cap \mathcal{G}_Q|}{|\mathcal{G}_Q|}\right\},\tau  \right)
\label{eq:score}\\
\text{where }CC(\mathcal{G^*})=\{\mathcal{G}_1^*, \ldots, \mathcal{G}_n^*\} \text{ and }\nonumber\\
max(S, \tau) =\{ max(S) \text{ if } max(S) > \tau, 0 \text{ otherwise} \}.\nonumber
\end{eqnarray}
The connected component that yields the highest score above the threshold $\tau$, together with its intersected edges, is identified as the matching subgraph corresponding to the query. 
The intersected edge-induced graph extracted from this connected component is returned as a response to the query.

\section{Results}

We evaluate our approach on four DARPA TC \cite{DARPA2014} datasets--Theia, Trace, Cadets, and FiveDirections--which feature eight distinct attack scenarios as described in~\cite{darpa-2018}. The Theia dataset was collected from hosts operating on Ubuntu 12.04, the Trace dataset was collected from hosts operating on Ubuntu 14.04, the Cadets dataset was obtained from a FreeBSD 11.0 host, and the FiveDirections dataset was collected from a Windows 7 machine. The attack scenarios used to evaluate our approach include an Nginx server backdoor, a Firefox backdoor, a backdoor with one of Firefox's extensions (password manager), and a phishing email with a malicious Excel document.

In this section, we start by evaluating the efficiency of our graph reduction strategies. 
Subsequently, we assess the capacity of order embeddings to represent subgraph relationships using DARPA TC datasets. 
Next, we examine our technique's ability to search for and identify subgraphs with two types of queries: those derived from converting DARPA TC attack logs into query graphs and those representing generic system activities. Finally, we compare \system to other hypothesis-driven threat hunting methods in terms of both effectiveness and performance.

\subsection{Reduction in Graph Size}
\label{sec:res-reduction}

\begin{table}[t]
    \caption{Reduction in Ego-Graph Size During Graph Creation Process (GS: Graph Simplification, DEM: Dependence Exploision Mitigation, BR: Behavior-Preserving Reduction)}
    \resizebox{\linewidth}{!}{
    \centering
    \begin{tabular}{c|c|c|c|c|c|c|c|c} 
        \toprule
        \multirow{2}{*}{Dataset} & \multicolumn{2}{c|}{Initial}&\multicolumn{2}{c|}{GS (Sec. \ref{subsec:graph_simplification})}&\multicolumn{2}{c|}{DEM (Sec. \ref{subsec:dependence_explosion})}&\multicolumn{2}{c}{BR (Sec. \ref{subsec:behavior_reduction})}  \\ \cline{2-9}        
                        & N & E         & N & E            & N & E     & N & E     \\\hline
        Theia           & 206k & 13M    & 22k & 6.9M        &159&336	&19&38      \\ 
        Trace           & 29k & 490k    & 400 & 1461        &329&1303	&16&28      \\ 
        Cadets          & 5k & 160k     & 1.9k & 76k        &43&75	    &12&15     \\ 
        FiveDirection   & 25k & 8.7M    & 2.1k & 4.2M       &350&1034   &71&527      \\ 
                
        \bottomrule
    \end{tabular}
    }
    \label{tab:effect}
\end{table}

We demonstrate the effectiveness of our graph creation method in terms of reduction in the graph size.
Table~\ref{tab:effect} summarizes the results obtained for each dataset, where we compute the average count of nodes and edges in $3$-hop ego-graphs of process nodes.
The first column of the table presents the average number of nodes and edges in each ego-graph after the graph simplification steps, up until the entity abstraction step described in Sec.~\ref{subsec:graph_simplification} is applied. (Since DeepHunter \ also applies these steps \cite{wei2021deephunter}, we consider this as our starting point). 
We then process the provenance graph by applying all remaining graph reduction steps.
Our results show a substantial reduction in the ego-graph size across all datasets. 
For instance, on the Theia dataset, the ego-graphs initially contain 206K nodes and 13M edges, while the final ego-graphs contain only 19 nodes and 38 edges, on average.
The variation across datasets can be attributed to the nature of graphs where the node degrees are much smaller in the Trace and Cadets dataset. 

These results demonstrate the effectiveness of our approach in reducing the size of ego-graphs while still preserving the diverse behaviors 
exhibited by processes. 
In fact, we observe that several reduced ego-graphs $\mathcal{G}_p$ are duplicated, containing identical nodes and edges.
To ensure all graph relationships are learned on an equal footing, regardless of their prevalence, we retain only one sample from each set of repeated ego-graphs.
This reduces the total number of ego graphs from 15k, 235k, 195k, and 17k to 1k, 11k, 3k, and 3k for Theia, Trace, Cadets, and Five Directions datasets, respectively.
\textcolor{black}{It's essential to clarify that several other approaches have been proposed to simplify provenance graphs for various computational purposes. (See Section \ref{sec:provg-simplification} for a brief review.) However, the effectiveness of a graph simplification method should be evaluated based on its intended primary task. For instance, a graph simplification technique designed to support forensic tractability or anomaly detection may not be suitable for our specific use case, and vice versa. In the context of searching for threat behaviors, it is worth mentioning that Poirot \cite{milajerdi2019poirot} does not employ graph reduction, whereas DeepHunter \cite{wei2021deephunter} suggests preserving parts of the graph around EDR alerts.}

\subsection{GNN Architecture and Parameter Selection}
\label{sec:grid_search}

In the design phase of our model, we evaluate the performance of the subgraph prediction function (Sec. \ref{subsec:subgraph}) on the Theia dataset to determine the most suitable GNN architecture and optimal model parameters.
We train the models using 80\% of the 3-hop ego-graphs and reserve 20\% for testing. During the training phase, we ensure  the test samples are not seen.
We conduct the training on 400 batches with a batch size of 1024, which includes an equal number of positive and negative target-query pairs. After the training, we evaluate the models on 10 batches.

To identify the most effective GNN architecture for our system, we assess the performance of several well-known graph neural network architectures, such as  GCN\cite{gcn:iclr:2017}, GIN~\cite{xu2018powerful}, and GraphSage~\cite{graphsage:nips:2017}.
Additionally, we experiment with the Multi-Relation GNN architecture~\cite{scarselli2008graph, rgcn2018}, where each edge type and direction are represented separately.
Following a thorough evaluation, we determine that the multi-relational GraphSage model, which integrates GraphSage with the Multi-Relation GNN, delivers the best performance among the tested architectures.

We also analyze the impact of the number of layers and aggregation method used to obtain subgraph embeddings on the model's performance. 
Although the performance differences are not substantial, using three layers yields the best results. 
In addition, we explore a variety of pooling methods, such as add pooling, mean pooling, graph multiset pooling~\cite{graphsetpooling}, and utilizing only the anchor node's embedding. Our findings indicate that add pooling, which aggregates the embeddings of all nodes in the graph, surpasses the other pooling techniques. We present the results of these tests in Appendix Sec. \ref{sec:app-grid} in Fig.~\ref{fig:grid-app}.
We conduct further experiments to identify optimal values for batch size, scheduling scheme, weight decay parameter, and embedding size. The results reveal that, apart from the embedding size, the choice of other parameters does not significantly affect the performance. We observe that improvements become marginal when the embedding size exceeds 256 dimensions. Consequently, we choose this value for our experiments.

\subsection{Power of Order Embeddings}
\label{subsec:powerorderembedding}

\begin{figure}[!ht]
\centering
   \includegraphics[width=0.6\columnwidth]{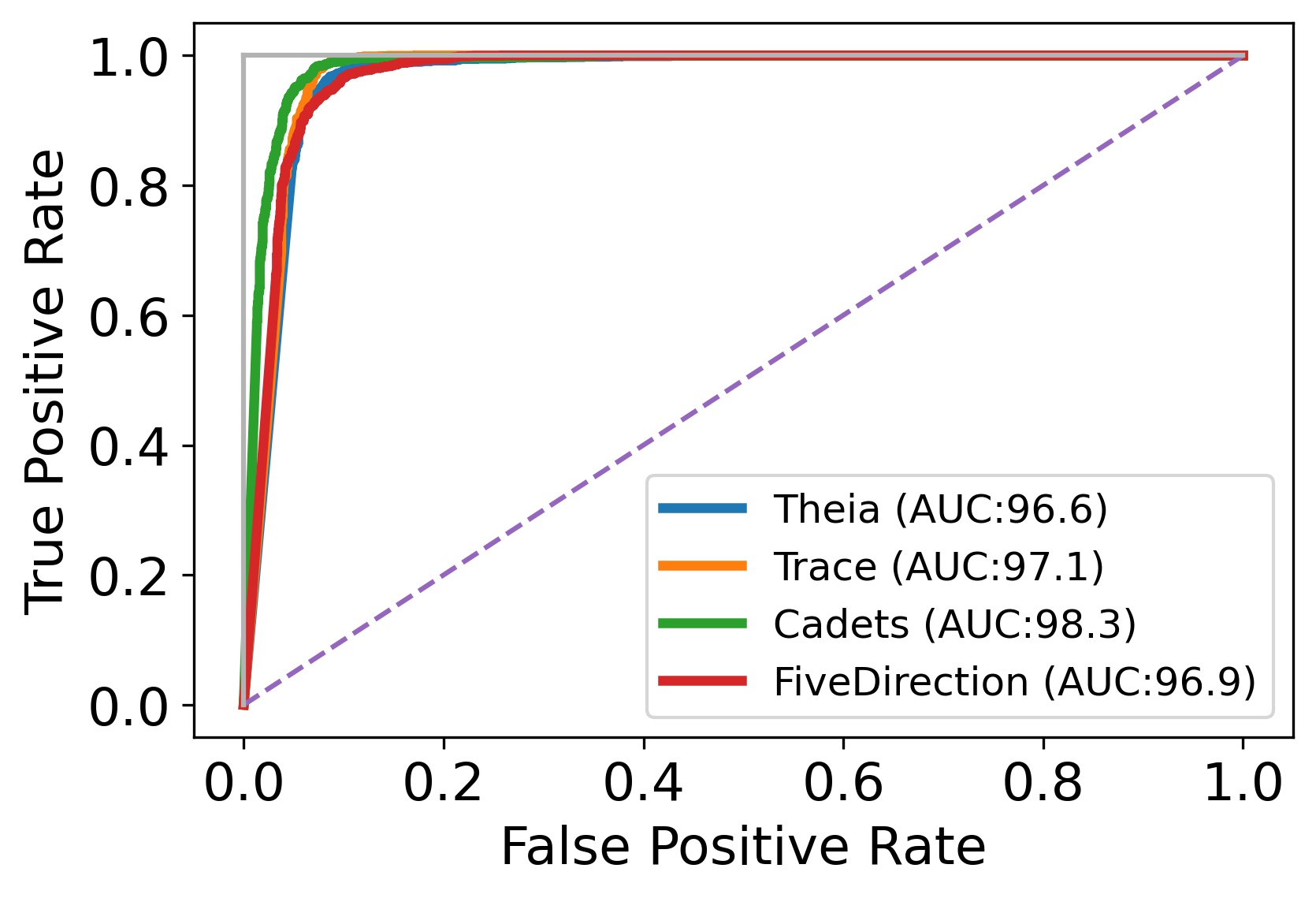}
   \caption{ROC curves for validating subgraph relationship between 3-hop graphs using our model on all the datasets}
  \label{fig:GNN_roc}
\end{figure}

We initially assess the effectiveness of order embeddings in identifying subgraph relationships between two graphs. 
We train a separate model for each dataset to learn the subgraph prediction function, i.e., $\varphi(z_p, z_q)$.
We employ a 3-layer multi-relational GraphSage GNN architecture with add pooling and an embedding size of 256.
The ability of order embeddings to detect subgraph relationships among 3-hop graphs is illustrated by the ROC curves shown in Fig.~\ref{fig:GNN_roc}(a). 
The results indicate that our method exhibits strong performance, with AUC scores ranging from 96.6 to 98.3 across the datasets, effectively distinguishing positive queries from negative ones in the provenance graph.

\textcolor{black}{To examine the robustness properties of order embeddings, we use the same set of test samples from the Theia dataset while considering two types of changes in attacker behaviors. In the first scenario, we examine the inclusion of additional activities by the attacker on the system, relative to the query graph. This behavior can be emulated by reducing the number of nodes and edges in the query graph. In the second scenario, we assume the attacker performs fewer actions on the system compared to the query, which is simulated by 
removing nodes and edges from the target graph.
The ROC curves obtained under both scenarios are shown in Fig. \ref{fig:app_GNN_roc} in Appendix Sec. \ref{sec:app-robustness}. The results indicate that, in scenario \#1 where nodes and edges are eliminated from the query graphs, Fig. \ref{fig:app_GNN_roc}(a), order embeddings achieve AUC scores of 93.3 and 90.2 when removing 15\% of the query nodes and 45\% of the query edges, respectively. A similar behavior is observed in scenario \#2, where nodes and edges are removed from the target graph, Fig. \ref{fig:app_GNN_roc}(b). Our technique achieves AUC scores of 92.9 and 93.0 when eliminating up to 15\% of the query nodes and 45\% of the query edges from the target graphs, respectively.
Based on these findings, we can conclude that our technique is capable of handling imprecise queries.}

\subsection{Accuracy in Subgraph Entailment Decision}

We now test the performance of \system in accurately determining whether a query graph is entailed within a provenance graph by computing the subgraph matching scores, Eq. (\ref{eq:score}), between the query and the target graphs.
For this, we use the models generated earlier to evaluate the subgraph relations among 3-hop graphs. We test their performance on two sets: (i) attack queries underlying DARPA-TC datasets~\cite{darpa-2018}, and (ii) a new test set comprising 5-hop ego-graphs involving generic behaviors extracted from the test portion of the Theia dataset.

\subsubsection{DARPA-TC Attack Queries}
\label{sec:att-queries}

The DARPA-TC dataset consists of eight attack scenarios, each consisting of up to three processes.
We began by evaluating the subgraph prediction function, which involves extracting and searching process-centric ego-graphs from the query graph within the corresponding provenance graphs. 
Table~\ref{tab:POSdist} displays the number of matching ego-graphs compared to the total number associated with each process in the provenance graph. 
For instance, process P1 has 846 instances within the provenance graph in the TRACE dataset's attack query. Our subgraph prediction function identifies only two, or 2/846, as matching candidates. Notably, no missed matches are observed in any of the test scenarios.
Upon analyzing the false matches, we find that all returned ego-graphs are connected, and on average, 78.9\% of all query nodes appear in those ego-graphs.
A comparison between the number of query nodes in correctly-matching and incorrectly-matching ego-graphs reveals that the former contain, on average, 57\% more query nodes.
This indicates that our subgraph matching function effectively localizes the query within the provenance graph.
We proceed to calculate the overall graph-matching scores for each scenario by first merging all the returned ego-graphs into a single graph, $\mathcal{G}^*$.
Following that, we compute the corresponding scores $g(\mathcal{G}^*, \mathcal{G}_Q)$, as described in Sec. \ref{subsec:matching}.
The resulting score values consistently exceeded 0.9, indicating a high degree of matching accuracy in all scenarios.
\textcolor{black}{To assess the false matching rate, we also tested each trained model by querying them with attack queries from other datasets. The results showed that none of these queries yielded a match, with the} highest observed matching score being around 0.2.

\begin{table}[t]
    \footnotesize
    \caption{Number of Matches Identified Per Query Ego-Graph}
    \centering
    \begin{tabular}{c|c|c|c|c} 
        \toprule
        Dataset & Query & P1        & P2        & P3        \\ \hline
        Theia   & Q1    & 2/846     & 1/1       & 1/1       \\ \hline
        \multirow{ 2}{*}{Trace} 
                & Q2    & 1/21023   & 1/1       & -         \\ 
                & Q3    & 1/239     & 1/2       & -         \\ \hline
        \multirow{ 3}{*}{Cadets} 
                & Q4    & 3/15      & 1/1       & -         \\ 
                & Q5    & 1/15      & 1/1       & -         \\ 
                & Q6    & 2/15      & 3/4       & 1/2       \\ \hline
        \multirow{ 2}{*}{Five Direction} 
                & Q7    & 6/724     & -         & -         \\ 
                & Q8    & 9/724     & 3/10      & 1/1       \\ \hline
                
        \bottomrule
    \end{tabular}
    \label{tab:POSdist}
\end{table}

\subsubsection{Generic Queries}
\label{sec:GED_res}

We also evaluate the performance of \system on a set of queries that include generic system behaviors.
\textcolor{black}{A generic behavior refers to any system activity that does not necessarily involve attack events. These behaviors can be obtained by randomly subsampling ego-graphs of process nodes from the system-level provenance graph. However, this approach would result in a sampling that mainly comprises the most prevalent system behaviors, leading to repeated queries with the same behaviors. 
To address this, we adopt the same approach used during model training to learn diverse system behaviors (Sec. \ref{subsec:sample_generation}).
For this purpose, we generate ten batches of test samples, each consisting of 5-hop ego-graphs for both positive and negative sample pairs (5K each).
Here, a positive sample includes an ego-graph with a matching subgraph, while a negative sample includes an ego-graph with a non-matching but similar graph. Appendix Sec. \ref{sec:app-query} presents examples of positive and negative sample pairs.}

We validate the subgraph relationship by calculating the order violation penalty, $\varphi(z_p, z_q)$, between the ego-graphs of the provenance graph and those in the query. During the evaluation, we set the threshold 
$\tau_{ovp}$ to the value that produces the highest accuracy in our experiments. \textcolor{black}{The optimum value for $\tau_{ovp}$ varies between 0.025 and 0.045 depending on the dataset as can be seen in Appendix \ref{sec:app-grid}, Fig. \ref{fig:grid-app}.)}
Next, we combine the matching ego-graphs to create $\mathcal{G}^*$ and calculate the final matching score, $g(\mathcal{G}^*, \mathcal{G}_Q)$.
The resulting ROC curve for subgraph matching scores is displayed in Fig.~\ref{fig:system_roc}(a) (the exact match curve). The overall performance in determining whether the query graph is contained within the target graph results in an AUC value of 99.8.
\textcolor{black}{We also conducted an assessment of \system's robustness to imprecise queries. Our findings, showcasing its robustness against node and edge removal, as well as its ability to match mutated malware behavior, are detailed in Appendix Sec. \ref{app:robust_generic}.}

\subsection{Comparison with Previous Approaches}
\label{sec:comparison}
This section presents a comparative evaluation of our technique against other hypothesis-driven threat hunting approaches, specifically  Poirot and DeepHunter~\cite{wei2021deephunter}.
As the source codes for these systems are not publicly accessible, we have implemented them to the best of our ability for the purpose of this comparison.
\textcolor{black}{To expand upon the tests for detecting attack queries (Sec. \ref{sec:att-queries}) and conduct a more comprehensive evaluation with a larger sample size, we conducted additional tests involving generic system behaviors, similar to Sec. \ref{sec:GED_res}. For this comparison, we generated 5K positive and 5K negative query-target pairs from each dataset and evaluated all techniques on the same data.}

\paragraph{Poirot~\cite{milajerdi2019poirot}:} Poirot is a non-learning-based search method that falls under the category of methods requiring the entire search computation to be executed at query time. However, this constraint makes Poirot less suitable for handling large datasets or supporting highly responsive applications. During operation, Poirot searches for all query nodes in the provenance graph until it finds the first acceptable alignment that exceeds a predefined threshold or exhaustively examines all possible paths.
As the graph size or number of queries increases, this extensive search also leads to a tendency to generate false alignments.
When evaluating the alignment of query nodes with those on a graph path, the Poirot algorithm calculates an attacker influence score by considering the number of compromised ancestor processes present on the path.
However, Poirot's definition of an ancestor process is not clearly established.
To address this ambiguity, we conduct experiments using various definitions of ancestor processes, including immediate ancestors, top-most ancestors, and top-most ancestors identified through clone, fork, or execute events. We report the best-obtained values.

We first assess the search complexity of Poirot on Theia and Trace datasets. As these datasets contain only a single instance of some processes in the queried attack behavior (i.e., mail and profile processes), they are quite favorable for Poirot.
The execution time of Poirot on these queries has been reported to vary from several minutes \cite{milajerdi2019poirot} to over an hour~\cite{threatraptor2021}, which can be attributed to the level of optimization in the code.
Upon analysis, we determined that the influence score computation requires traversing a large portion of the graph. For instance, the query on the Theia dataset for the Firefox backdoor vulnerability traverses around 40\% of all edges and 13\% of all nodes. 
This factor can significantly hinder Poirot's performance, particularly with larger and more complex datasets.

\textcolor{black}{
The performance of Poirot in searching generic system behaviors entailed in the test samples associated with the three datasets are presented in Table~\ref{tab:combined_table}\eat{ (lines 4, 9, and 13)}.
These results show that \system and Poirot perform very similarly in correctly matching query system behaviors entailed in positive samples to corresponding target graphs, with \systemNoSpace slightly outperforming Poirot by around 1\% in all cases.
However, as shown in the last column of Table~\ref{tab:combined_table}, Poirot produces a considerably higher false-positive rate on Theia (3\%) and Trace (5\%) datasets, compared to \systemNoSpace, 
which falsely matches only around 0.01\% of the neqative samples. 
On the Cadets dataset, Poirot also yields a low false-positive rate (0.69\%), which can be attributed to the relatively simpler nature of the graphs in the Cadets dataset (Table \ref{tab:effect}).
It is important to note that both \system and Poirot significantly outperformed other techniques used for testing.
}

\paragraph{IsoRankN~\cite{liao2009isorankn}} IsoRankN is a network alignment method designed to identify equivalent nodes or edges in multiple networks. It achieves this by optimizing a global objective function that measures network similarity through node alignment. The method utilizes a spectral clustering algorithm to break down the networks into smaller subgraphs and aligns them based on their topological properties.
\textcolor{black}{However, the results\eat{(lines 1, 6, and 11)} indicate that this method is not highly effective for the subgraph matching task.}

\paragraph{DeepHunter \cite{wei2021deephunter}:} The subgraph matching technique most similar to ours is DeepHunter, which utilizes an NTN to learn the subgraph relationship between graph embeddings.
DeepHunter relies on EDR alerts to reduce the provenance graph by identifying query seed nodes close to IoCs and to minimize falsely matching behaviors.
However, as we don't have access to graphs created around the IoCs, we ran it on graphs created using our method.
To compare with DeepHunter, we integrated its search component with our graph creation module and fine-tuned its parameters for each dataset. 
We trained the model until its performance plateaued and conducted the tests accordingly.
The corresponding results are presented in Table \ref{tab:combined_table}\eat{ (lines 3, 8, and 13)}.
\textcolor{black}{These results clearly illustrate that \system surpasses DeepHunter by approximately 15-25\% in terms of accuracy.}

\paragraph{SimGNN \cite{bai2019simgnn}:} SimGNN uses a more advanced architecture than DeepHunter that also incorporates node-alignment into its decision.
\textcolor{black}{For our tests, we optimized its parameters for each dataset}.
Our results show that, despite this additional modeling step, SimGNN performs similar to DeepHunter\eat{ (lines 2, 7 and 12)}, with 
SimGNN performing slightly worse, consistent with the findings reported in \cite{bai2019simgnn} .
DeepHunter slightly outperforming it outperforming in on Theia dataset by  consistent with the findings reported by DeepHunter.

\paragraph{\textcolor{black}{Runtime Analysis:}}
\textcolor{black}{
Our technique's offline phase incurs a fixed one-time cost for each dataset, involving the creation of training samples and model training. 
In our evaluation, we determined that the collective steps of the offline phase take roughly 400, 287, and 250 minutes for Theia, Trace, and Cadets datasets, respectively.
As part of the offline stage, the trained model is also employed to precompute subgraph embeddings for all ego-graphs within the provenance graph.
This process averages around 11 minutes for each dataset.
The online phase involves generating embeddings for query graphs and performing searches for matching subgraphs using precomputed embeddings.
Across all datasets, our technique completes processing 10,000 samples in approximately 48 seconds. In contrast, Poirot's execution times for the same set of queries are considerably longer, requiring 1166, 392, and 327 minutes for Theia, Trace, and Cadets datasets, respectively.
}

\begin{table}[!ht]
\centering
\caption{\textcolor{black}{Performance Comparison on Different Datasets}\eat{\textcolor{blue}{Tablo'ya satir numarlari da ekleyebiliriz, ama sart degil.}}}
\resizebox{\linewidth}{!}{
\begin{tabular}{|l|l|c|c|c|c|c|c|}
    \hline
    \multirow{2}{*}{\rotatebox[origin=c]{90}{}} & \multirow{2}{*}{Method} & \multicolumn{6}{c|}{Metrics} \\ 
    \cline{3-8} 
    & & Acc. & F1 & AUC & Prec. & Recall & FPR \\
    \hline
    \multirow{5}{*}{\rotatebox[origin=c]{90}{Theia}} & IsoRankN & 63.20 & 62.85 & 63.06 & 63.46 & 62.26 & 35.84 \\
     & SimGNN & 83.28&84.49&90.39&78.77&91.11&24.56 \\
     & DeepHunter & 83.67&84.42&90.86&80.69&88.53&21.19 \\
     & Poirot & 97.38 & 97.44 & 97.46 & 95.16 & \textbf{99.84} & 5.07  \\
     & \system & \textbf{99.83}  & \textbf{99.84} & \textbf{99.81} & \textbf{99.98}  & 99.69 & \textbf{0.02}  \\
    \hline
    \multirow{5}{*}{\rotatebox[origin=c]{90}{Trace}} & IsoRankN  & 56.47 & 36.46 & 55.07 & 68.36 & 24.86 & 11.61 \\
     & SimGNN & 75.93&78.63&84.33&70.69&88.57&36.72 \\
     & DeepHunter & 74.93&77.45&83.51&70.36&86.13&36.28\\
     & Poirot & 97.99 & 98.01 & 98.40 & 97.03 & \textbf{99.02} & 3.02 \\
     & \system & \textbf{99.34} & \textbf{99.33} & \textbf{99.34} & \textbf{99.98} & 98.69 & \textbf{0.01} \\
    \hline
    \multirow{5}{*}{\rotatebox[origin=c]{90}{Cadets}} & IsoRankN & 62.17 & 44.15 & 56.96 & 84.35 & 29.90 & 5.54\\
     & SimGNN & 84.50&85.55&90.27&80.10&91.80&22.80 \\
     & DeepHunter & 84.11&85.21&90.43&79.69&91.55&23.34\\
     & Poirot & 98.18 & 98.16 & 99.66 &  99.30 & 97.05 & 0.68 \\
     & \system & \textbf{99.78} & \textbf{99.76} & \textbf{99.80} & \textbf{99.96} & \textbf{99.61} & \textbf{0.03}\\
    \hline
\end{tabular}
}
\label{tab:combined_table}
\end{table}

\subsection{\textcolor{black}{Ablation Study}}
\textcolor{black}{
To assess the impact of various subcomponents of \system on its performance, we conducted an ablation study.
In this regard, the graph creation component has three processing steps that are indispensable for the operation of our technique. 
These include the graph simplification (GS), dependence explosion mitigation (DEM), and the graph partitioning (GP) steps.
The GS and DEM steps play a crucial role in reducing the size of the graphs, making them suitable for graph representation learning approaches. 
On the other hand, the GP step is responsible for generating the required ego-graphs, facilitating the subsequent operations of our system.
As shown in Table \ref{tab:effect}, the size of ego-graphs is unmanageably large for all datasets, except for the Trace dataset, before the application of the DEM step. 
Consequently, the ablation study involving the DEM step, in addition to the behavior-preserving reduction (BR) step, is limited to the Trace dataset.
In our study, we also evaluate the impact of the subgraph matching (SM) step, which allows \system to process large query graphs by partitioning them into $k$-hop ego-graphs.
}

\textcolor{black}{
As part of our study, we repeated the experiments in Sec. \ref{subsec:powerorderembedding} and \ref{sec:GED_res} by partitioning the provenance graph into both 3-hop (lines 1-3) and 5-hop (lines 4-7) ego-graphs when generating subgraph representations. 
The query graphs, respectively, include 3-hop and 5-hop ego-graphs in both cases. 
Results corresponding to different combination of subcomponents are presented in Table \ref{tab:ablation}.
Specifically, the DEM step alone impacts the accuracy by 1.5\% and 5.6\% (lines 1-2 and 5-6) depending on the size of ego-graphs.
Moreover, the results demonstrate that the BR step leads to a substantial increase in accuracy, around 3\% (lines 2-3 and 5-6), considering both $3$- and $5$-hop ego-graphs.
The effectiveness of the SM step is demonstrated in lines 5 and 7. In line 5, where the query and ego-graphs have the same size, the SM function is not required. 
However, in line 7, the query graphs are partitioned into 3-hop ego-graphs before the match is evaluated using the SM function. 
Consequently, the SM step enables the utilization of smaller graph representations (3-hop ego-graphs instead of 5-hop ego-graphs), resulting in a more precise query matching. This improvement boosted the accuracy by almost 5.5\%, increasing it from 93.33\% to 98.83\%.
When combined together, the BR and SM steps yield an additional improvement of 0.5\%, pushing the overall accuracy to an impressive 99.34\%.
}
\begin{table}[!ht]
\centering
\caption{\textcolor{black}{Ablation Study (GS: Graph Simplification, DEM: Dependence Exploision Mitigation, GP: Graph Partitioning BR: Behavior-Preserving Reduction, SM: Subgraph Matching)}}
\resizebox{\linewidth}{!}{
\begin{tabular}{l|l|c|c|c|c|}
    \cline{2-6} 
     &\multirow{2}{*}{System Components} & \multicolumn{4}{c|}{Metrics} \\ 
    \cline{3-6} 
    && Accuracy & Precision & Recall & FPR \\
    \cline{2-6} 
    1&GS+GP (k=3) &90.01 & 87.37 & 93.55 & 13.52 \\  
    2&GS+DEM+GP (k=3) & 91.52 & 92.96 & 90.36 & 7.23 \\   
    3&GS+DEM+GP+BR (k=3) & 95.09 & 97.46 & 93.05 & 2.66 \\ \cline{2-6}  
    4&GS+GP (k=5) & 87.71 & 83.95 & 93.26 & 17.82 \\  
    5&GS+DEM+GP (k=5) & 93.33  & 97.21 & 90.21 & 3.01 \\  
    6&GS+DEM+GP+BR (k=5) & 96.06 & 93.51 & 99.00 & 6.87 \\ \cline{2-6}  
    7&GS+DEM+GP+SM (k=5)  & 98.83 & 98.67 & \textbf{99.79} & 1.31 \\  
    8&GS+DEM+GP+BR+SM (k=5)  & \textbf{99.34}  & \textbf{99.98}  & 99.69 & \textbf{0.02}  \\
    \cline{2-6} 
\end{tabular}
}
\label{tab:ablation}
\end{table}

\section{Related Work}
\subsection{Threat Hunting and Provenance Analysis}
Threat hunting is a proactive defense approach where experts continuously search for traces of an unknown attack.
Provenance graphs have been utilized in two main ways to expand existing threat hunting capabilities. 
The first one aims at leveraging the contextual information revealed by a graph representation to discover irregularities and anomalies that may suggest
malicious activity.
To this objective, one group of work applied statistics-based techniques to identify unlikely events or event chains \cite{liu2018towards,hassan2019nodoze,wang2020you}.
Another group of work proposed clustering and learning-based techniques to distinguish between benign and anomalous patterns in the provenance graph.
These works applied several intuitions to graph analysis such as the use of 
graph sketching techniques \cite{manzoor2016fast, han2020unicorn}, 
graph embedding techniques \cite{li2021hierarchical, wang2022threatrace},
knowledge graph embedding techniques \cite{zeng2021watson, zengy2022shadewatcher}, and 
sequence-based neural embedding methods \cite{liu2019log2vec, wang2020you, alsaheel2021atlas}.

The other use of provenance graphs for threat hunting relates to hypothesis-driven investigations. 
When knowledge about a new threat is obtained from threat intelligence sources and feeds, offering intelligence on the most current tactics, techniques, and procedures utilized by attackers \cite{mitre-attack}, threat hunters will search within their own environment for those specific attack behaviors.  
In the context of provenance graphs, this search task can be formulated as a graph pattern matching problem where 
an observed attacker behavior is expressed as a query graph, and its entailment within a provenance graph has to be determined.
Given the NP-completeness of exact graph matching, this formulation of the problem calls for the use of approximate methods. 
Most inexact methods rely on heuristics to select appropriate seed nodes, and then expand to neighboring nodes according to predetermined rules to match the topology as well as node and edge features \cite{tian2007saga, tong2007fast, khan2013nema, pienta2014mage, liu2019g, milajerdi2019poirot}. 
To perform this search rapidly, Poirot examines all paths from the seed node to other nodes and retains only those that are more likely to be under the influence of an attacker for alignment by ensuring that the processes along a selected path share a common ancestor in their process tree.  
With a similar motivation, DeepHunter generates graph embeddings to evaluate the alignment of query graph with subgraphs extracted from the provenance graph   
and uses a neural tensor network to model the relation between two graph-level embeddings. 
To reduce the complexity of the search, the provenance graph is reduced by only considering subgraphs extracted around nodes related to suspicious events identified by alerts of an EDR tool.

\vspace{-0.2cm}
\subsection{Provenance Graph Simplification}
\label{sec:provg-simplification}

The substantial size of provenance graphs presents a significant challenge for conducting timely analyses, thus impeding their practical usability. 
To address this issue, one strategy is to utilize the limited memory resource efficiently trough adopting compact representations for referencing nodes and edges 
\cite{hossain2017sleuth, hossain2018dependence}, identifying the relevant portion of the graph for analysis and caching only that part in main memory \cite{hassan2020we}, or losslessly compressing the graph \cite{fei2021seal}.
While these approaches preserve the utility of data, their effectiveness inevitably diminishes as the complexity of the graphs increases.

Therefore, to simplify provenance graphs, several works have proposed preprocessing methods by removing logs that are associated with
dead system entities \cite{lee2013loggc}, repeated events \cite{xu2016high}, 
and events comprising nodes that are disconnected from the backward and forward tracing graph of a symptomatic, point-of-interest event \cite{alsaheel2021atlas} or (EDR) threat alert event \cite{hassan2020tactical}.
While lossy reduction techniques cannot provide a foolproof guarantee that every analysis task will yield the desired outcome, it's possible to perform the reduction in a way that doesn't compromise the objective of the forensics analysis. 
To achieve this objective, three reduction techniques have been proposed. The first, causality-preserving reduction, aims to eliminate events that are redundant for causality reasoning \cite{xu2016high}. 
The second technique, dependency-preserving reduction, reduces causality to reachability and removes events (i.e., edges of the graph) more aggressively, provided that system entities required for backward and forward tracing queries can be correctly identified \cite{hossain2018dependence}. 
The third technique, attack-preserving reduction, has been proposed to preserve attack-relevant causal relations while eliminating those related to benign process activities \cite{michael2020forensic,fang2022back}.
In contrast, when searching and matching graph patterns in provenance graphs, it is crucial to capture a wide range of local relations, regardless of their frequency of occurrence. Thus, log reduction techniques that aim to preserve causality or dependency relations along traces in a graph are not ideally suited for learning subgraph relationships.

Another key challenge in analyzing provenance graphs is to mitigate the risk of false dependencies.
As the number of nodes and edges in the graph increases, the potential dependencies between system entities grow exponentially, giving rise to the dependency explosion problem. 
The impact of this explosion is more severe for (long-running) high fanout processes and frequently accessed files.
Therefore, many works proposed to address the issue of dependence explosion through execution partitioning of graphs~\cite{lee2013high, ma2016protracer, ma2017mpi, kwon2018mci}, tag propagation for information flow tracking \cite{hossain2020combating}, and incorporation of system and application logs \cite{hassan2020omegalog}. 
An alternative and less costly approach to determining true dependencies between system entities is to create versioned graphs.
Versioning allows encoding the time order of events by creating a version of a process or file when its state changes. 
Several methods have been proposed to reduce the number of node versions and edges while preserving dependencies \cite{bates2015trustworthy, pasquier2017practical, hossain2018dependence}.

\subsection{\textcolor{black}{Graph Neural Networks and Subgraph Matching}}
Inspired by the success of graph learning methods in several prediction tasks, several graph learning-based methods have been recently proposed to solve approximate subgraph matching task.
These methods use graph neural networks to generate node level embeddings that encode the neighborhood structures and features of nodes as well as edges.
Then, resulting embeddings are used to model the relation between the data and query graphs.
When evaluating the match between two graphs, a common approach is using graph embedding models \cite{lou2020neural, bai2019simgnn, davitkova2021lmkg}.
This involves learning an inductive function that embeds graphs into a vector space such that similar graphs are mapped closely while dissimilar ones are far apart. 
At the graph embedding stage, the individual node embeddings are pooled together through one of several schemes as the final graph-level representation \cite{xu2018powerful}.
One challenge these models face is reflecting minute structural differences between graphs in the resulting graph-level embeddings. 
To address this, the larger data graph is partitioned into subgraphs to better emphasize node-level information, and 
the evaluation is done between the query graph and those smaller subgraphs comprising the data graph. 

An alternative approach to graph embedding is to allow a model to incorporate more complex relation information instead of independently mapping each graph to a vector. 
This is realized by jointly computing a similarity score between a pair of graphs. 
With the cross-graph matching approach, the reasoning of the relation between two graphs is made by modeling the node-to-node interactions \cite{roy2022interpretable, lan2021sub,li2019graph, wu2019relation, xu2019cross} or by modeling the graph-to-graph interactions \cite{bai2019simgnn}. 
Since the graph-learning model attends a pair of graphs jointly, cross-graph matching methods are potentially stronger than the graph embedding models, and they can be made more resistant to slight variations between graphs. 
This gain, however, comes at the cost of increased computational complexity.
Because the similarity computation has to be done in an online setting (i.e., after the query graph is presented) where 
the number of pairs of query and target subgraphs that need to be compared depends on the size of the data graph.
Therefore, for large-scale graph instances,  cross-graph matching methods are not feasible.
In contrast, graph embedding models allow operating in a batch setting by precomputing the embeddings for data graphs, thereby
limiting the online computation step to computing a similarity measure between a pair of embeddings. 
In the context of threat hunting, both the size of the provenance graphs and the potential number of queries are considerably large. Consequently, a graph embedding approach emerges as a more suitable solution to perform search in provenance graphs.

\section{Limitations and Conclusions}
The design of \system emphasizes learning diverse system behaviors, rather than focusing on \textcolor{black}{the} prevalence of those behaviors within the system. 
This approach can lead to ambiguous representations for queries involving repetitive activities.
For instance, a malicious software transferring or encrypting large number of files under a directory may be represented by only a few write events to different file object abstraction categories, making it harder to discern the underlying pattern.
A potential solution to address such cases is to train a new model tuned to learn frequently exhibited behaviors. Queries involving such repetitive behaviors can then be searched using this specialized model.

For efficiency and accuracy in matching, our technique performs search on reduced provenance graphs. 
Part of this reduction involves abstracting system entities by assigning them higher-level category labels. 
Although this abstraction based on entity's function enhances matching capability, the loss of specificity (such as process and file names, IP addresses) may contribute to false matches.
Results from real-life queries describing DARPA TC attack behaviors demonstrate that our technique can accurately match abstracted entities on unreduced provenance graphs.
However, for much larger graphs with increasing number of queries, this may lead to additional false matches, such as when two queries represent different behaviors but map to the same abstract behavior representation. 
Our approach can be adapted to this search mode by maintaining additional indexing information alongside the precomputed and stored ego-graph embeddings. 
By including information on the anchor process of an ego-graph, we can limit the evaluation of subgraph relationship only to matching entities.

One area that may offer further improvement to our technique involves utilizing deeper GNNs to better leverage their increased expressive power. 
Our initial tests reveal that using three or four layers yields similar performance. 
Using higher number of layers requires handling larger graphs during the graph creation phase and reducing batch sizes to accommodate target and query graph pairs within GPU memory during training. 
Exploring deeper networks will be considered in our future work.
Additionally, ensuring the generalizability of our technique to support search across different systems requires an operating system-agnostic representation for system entities and edge types. This aspect will also be considered in our future work.

\bibliographystyle{ACM-Reference-Format}
\bibliography{main-refs,software}

\appendix
\section{Characteristics of Graph Datasets}
\label{sec:graphDatasets}

\textcolor{black}{Table \ref{tab:graph_size} showcases the average number of nodes and edges for various graph datasets employed in
different application contexts. These datasets encompass biology (Mutag-MT, Predictive Toxicology Challenge-PTC, Enzymes-EZ), chemistry (COX2), and image processing (MSRC\_21), alongside the characteristics of the provenance graph in the Theia dataset.
As can be seen, the number of nodes and edges in the Theia dataset is several orders of magnitude higher compared to other datasets.
}

\begin{table}[H]
    \caption{\textcolor{black}{Graph Dataset Statistics in Subgraph Matching Applications (Average Node and Edge Counts)}}
    \resizebox{\linewidth}{!}{
    \centering
    \begin{tabular}{c|c|c|c|c|c|c} 
        \toprule
             & MT & PTC & EZ & COX2 & MSRC\_21 & Theia  \\ \hline
        \# Nodes & 18 & 14  & 33 & 41   &  77     & 206K               \\
        \# Edges & 39 & 15  & 124& 216  &   198   & 13M                \\
        \bottomrule
    \end{tabular}
    }
    \label{tab:graph_size}
\end{table}

\section{Abstraction Categories}
\label{sec:abstract_category}

\begin{table}[!ht]
\centering
\caption{Process and File Abstraction Categories for Linux Operating System}
\begin{tabular}{c}
\hline
bin, cache, com, data, dbus-vfs-daemon, dev, devd, \\
digit, dns, etc, home, lib, lib64, man, other, \\
proc, root, run, sbin, stream, sys, tmp, unknown, \\
usr, usrbin, var, vi, www  \\ \hline
\end{tabular}
\label{tab:abstarctions_linux}
\end{table}

\begin{table}[!ht]
\centering
\caption{Process and File Abstraction Categories for Windows Operating System}
\begin{tabular}{c}
\hline
c:program files, c:program files (x86), c:programdata, c:users, \\
c:windows, device, program files, program files (x86), \\
programdata, registry, systemroot, users, \\
windows, c:deploy-keys, c:lwabeat, c:program, \\
c:program files, c:program files (x86), c:programdata, c:recovery, \\
c:system volume information, c:tcssl, c:users, c:windows, \\
d:extend, d:recycle.bin, d:system volume information \\ \hline
\end{tabular}
\label{tab:abstarctions_linux}
\end{table}

\begin{table}[!ht]
\centering
\caption{Abstraction Categories for Network Objects}
\resizebox{\linewidth}{!}{
\begin{tabular}{c}
\hline
inter\_private\_inter, user\_local\_user, user\_private\_user, \\
user\_private\_reserved, user\_public\_inter, user\_public\_user, \\
user\_public\_reserved, reserved\_local\_user, reserved\_local\_reserved, \\
reserved\_private\_user, reserved\_private\_reserved, reserved\_public\_user \\ \hline
\end{tabular}
}
\label{tab:abstarctions_linux}
\end{table}

\section{Ego-graph Extraction Algorithm}
\label{sec:A1}

This algorithm is used to extract all ego-graphs of process nodes, i.e., $\mathcal{G}_{p}$, from the provenance graph $\mathcal{G}_\textbf{P}$.
The algorithm aggregates each versioned node's forward and backward neighbors starting from $0$-hop distance and extending to neighbors at $k$-hop distance, where the $0$-hop neighbor refers to the node itself. 
The $l$-hop neighbors of a node are aggregated from $(l-1)$-hop neighbors of the corresponding node's neighbors, except for versioned neighbors where different versions of the same node are considered to be at the same depth.
It must be noted that, for forward neighbors, the node with the next version, and for backward neighbors, the node with the previous version, are the only neighbors that can be reachable.
Additionally, we only compute the $0$-hop neighbors of object nodes with 0 in- or out-degree, which are added to sink nodes since they can only be reached in 1-hop.

Here, the function $In(n)$ returns all incoming neighbors of node $n$, while $Out(n)$ is used to obtain the outgoing neighbors. 
For a versioned node $n$ with version $i$, we aggregate the neighbors of node $n_{i+1}$ to calculate its forward neighbors and the neighbors of node $n_{i-1}$ to determine its backward neighbors.
The functions  $get\_next\_version(n)$ and $get\_prev\_version(n)$ are used within the algorithm to retrieve the next and previous versions of node $n$, respectively.

\begin{algorithm}[!tb]
\caption{Dynamic algorithm for provenance graph partitioning.}
\label{alg:partitioning}
\begin{algorithmic}[1]
\REQUIRE $\mathcal{G}_{P}$: reduced provenance graph, $k$: ego-graph hop count, $S$: set of sink nodes, $\mathcal{V}_P$: process nodes
\ENSURE $\mathcal{G}_{p}$: $k$-hop ego-graphs

\FORALL{$p\in \mathcal{V}$}
    \STATE $neigh[p]['forward']   \leftarrow p$
    \STATE $neigh[p]['backward']  \leftarrow p$
\ENDFOR

\FOR{$i=1,..,k$}
    \FORALL{$n\in \mathcal{V}$}
        \IF{ $n \in S$}
            \STATE $neigh[n]['forward'][0]   \leftarrow p$
            \STATE $neigh[n]['backward'][0]  \leftarrow p$
        \ELSE
        \STATE $ $\algorithmiccomment{Calculate forward neighbours}
        \FORALL{$w\in In(n)$}
            \STATE $neigh[n]['forw'][i] +\leftarrow neigh[w]['forw'][i-1]$
        \ENDFOR
        \STATE $nn = get\_next\_version(n)$
        \STATE $neigh[n]['forw'][i] +\leftarrow neigh[nn]['forw'][i]$
        
        \STATE $ $\algorithmiccomment{Calculate backward neighbours}
        \FORALL{$w\in Out(n)$}
            \STATE $neigh[n]['back'][i] +\leftarrow neigh[w]['back'][i-1]$
        \ENDFOR
        \STATE $pn = get\_prev\_version(n)$
        \STATE $neigh[n]['back'][i] +\leftarrow neigh[pn]['back'][i]$
        \ENDIF
    \ENDFOR
\ENDFOR
\FORALL{$p\in \mathcal{V}_P$}
    \STATE $\mathcal{G}_{p} \leftarrow neigh[p]['forw'][k] + neigh[p]['back'][k] $
\ENDFOR

\end{algorithmic}
\end{algorithm}

\section{Grid Search}
\label{sec:app-grid}

The performance impact of changing the GNN architectures, number of layers, \textcolor{black}{embedding dimension}, and aggregation method are evaluated on Theia, \textcolor{black}{Trace, and Cadets} datasets. 
The results of our grid-search analysis for all datasets are presented in Fig. \ref{fig:grid-app}.
\textcolor{black}{
Subsequently, upon fixing the GNN hyperparameter values accordingly, we investigated the influence of the threshold value $\tau$, which determines the maximally tolerable order violation penalty. 
Fig. \ref{fig:grid-app} presents the change in performance concerning the correct validation of the subgraph relationship as a function of $\tau$. 
}

\begin{figure}[H]
    \centering
    \includegraphics[width=1\columnwidth]{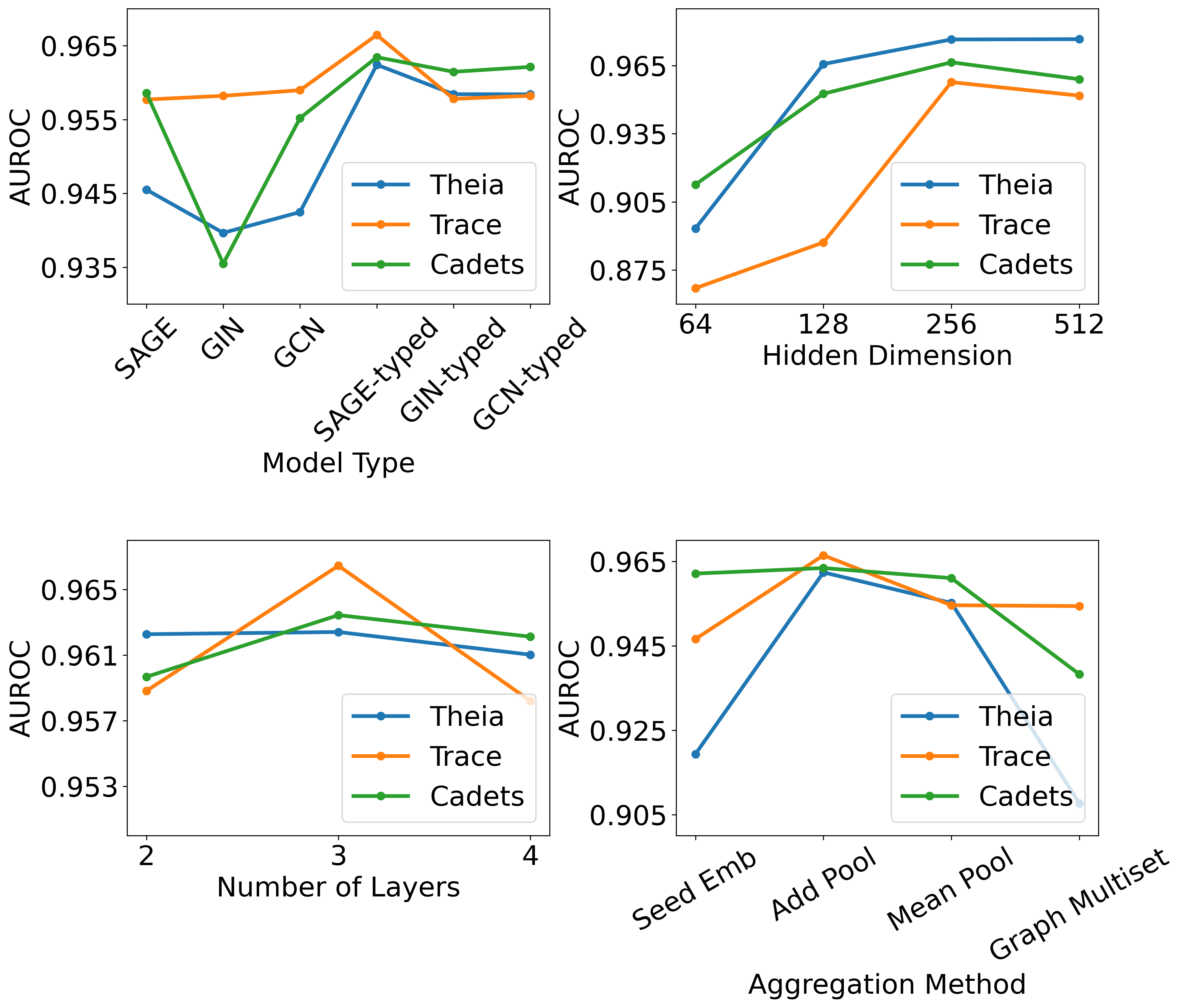}
    \caption{Impact of various GNN architectures, \textcolor{black}{embedding dimensions,} number of layers,  and aggregation methods on performance \textcolor{black}{across three datasets.}}
    \label{fig:grid-app}
\end{figure}

\begin{figure}[H]
    \centering
    \includegraphics[width=0.6\columnwidth]{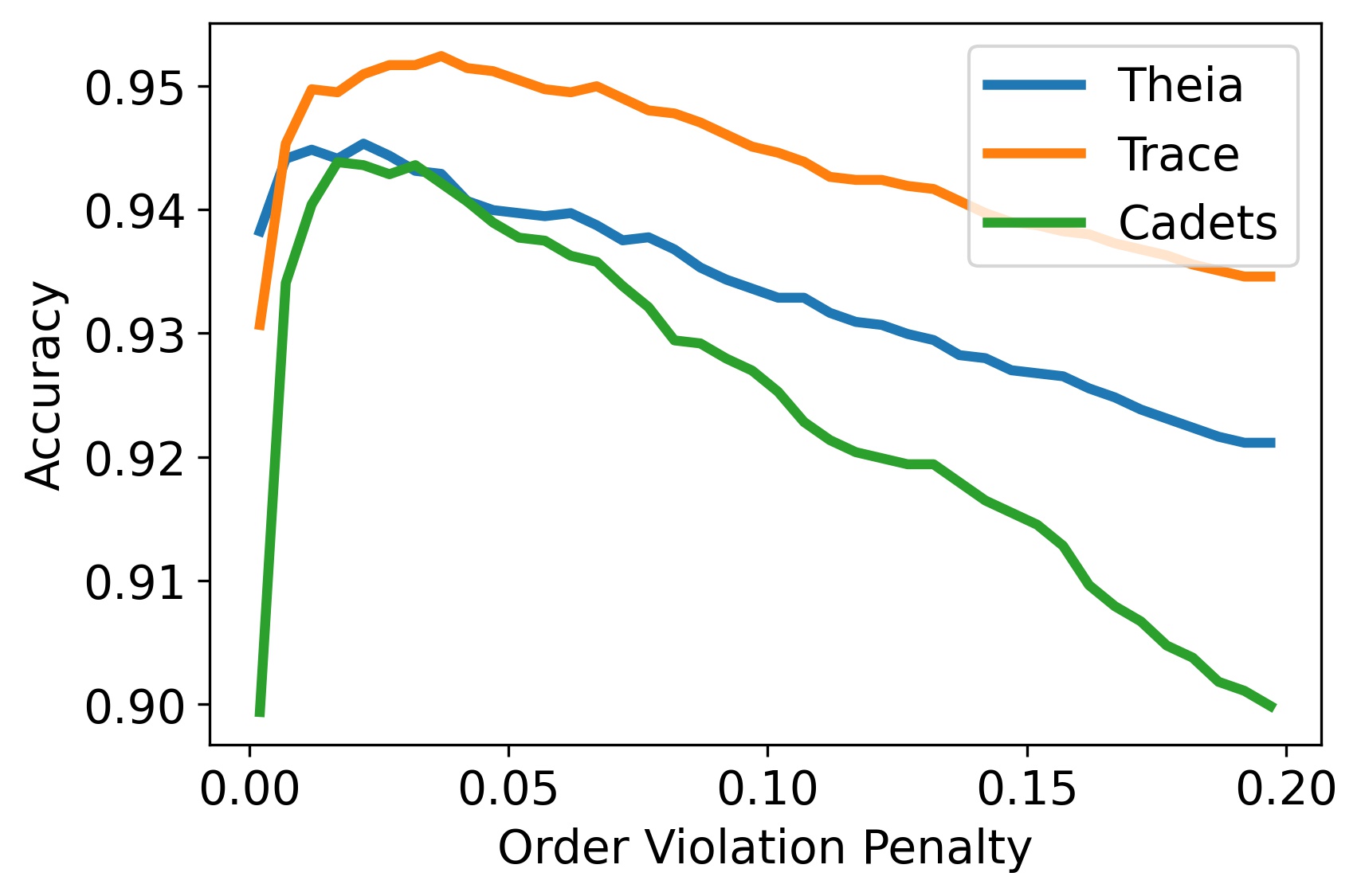}
    \caption{\textcolor{black}{Impact of order violation penalty values on subgraph relationship detection across three datasets.}}
    \label{fig:grid-app}
\end{figure}

\section{Additional Results on Robustness of Order Embeddings}
\label{sec:app-robustness}
\begin{figure}[H]
\centering
  \begin{minipage}[a]{0.48\columnwidth}
    \centering
    \includegraphics[width=0.9\columnwidth]{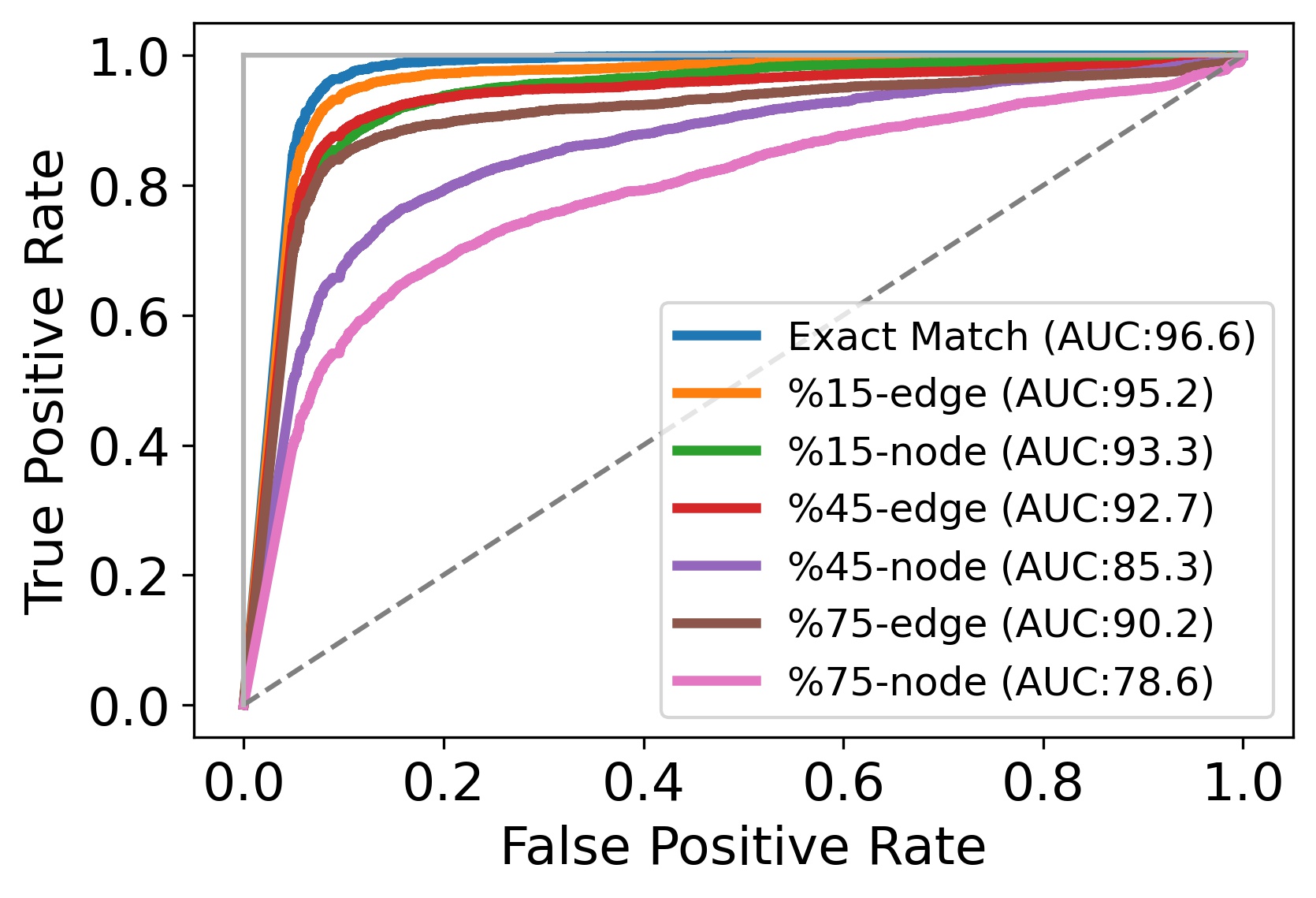} 
    \centerline{(a)}
  \end{minipage}
  \begin{minipage}[a]{0.48\columnwidth}
    \centering
    \includegraphics[width=0.9\columnwidth]{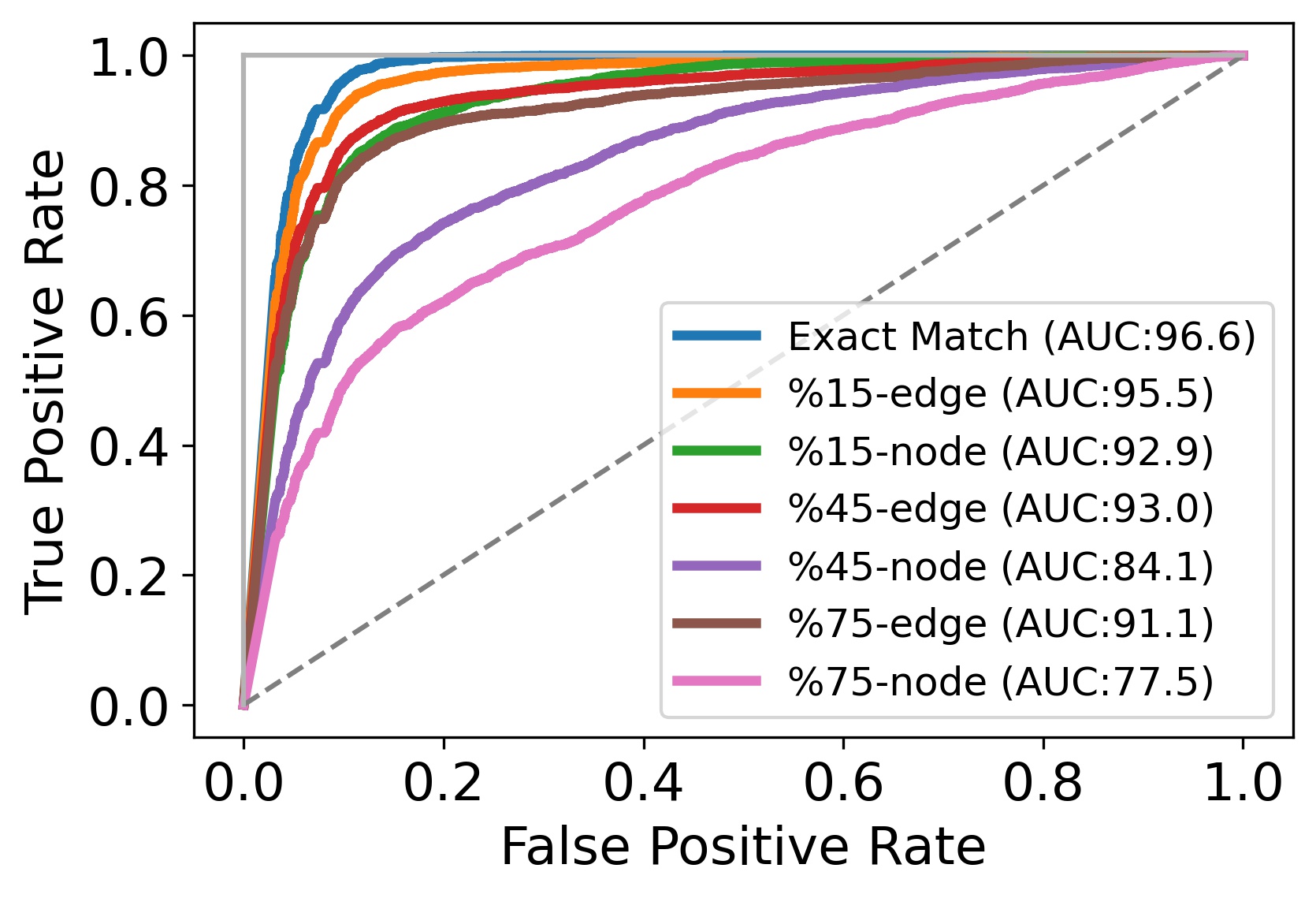}
    \centerline{(b)}
  \end{minipage}
   \caption{\textcolor{black}{ROC curves for validating subgraph relationship on the Theia dataset considering 3-hop graphs when a random portion of nodes
   and edges are removed from the (a) query graph (scenario \#1)} and (b) target graph (scenario \#2).}   
  \label{fig:app_GNN_roc}
\end{figure}

\textcolor{black}{We evaluate the robustness of order embeddings to the addition and deletion of nodes and edges.
Figure \ref{fig:app_GNN_roc}(a) shows ROC curves resulting from trimming the query graph, depicting the scenario where the attacker introduces extra attack actions into their known behavior.
Similarly, in Fig. \ref{fig:app_GNN_roc}(b), ROC curves are presented for the target graph's case,
reflecting the scenario where the attacker removes specific attack actions from their known behavior.}

\begin{figure*}
\centering
      \begin{minipage}[a]{1\columnwidth}
        \centering
        \includegraphics[width=0.9\columnwidth]{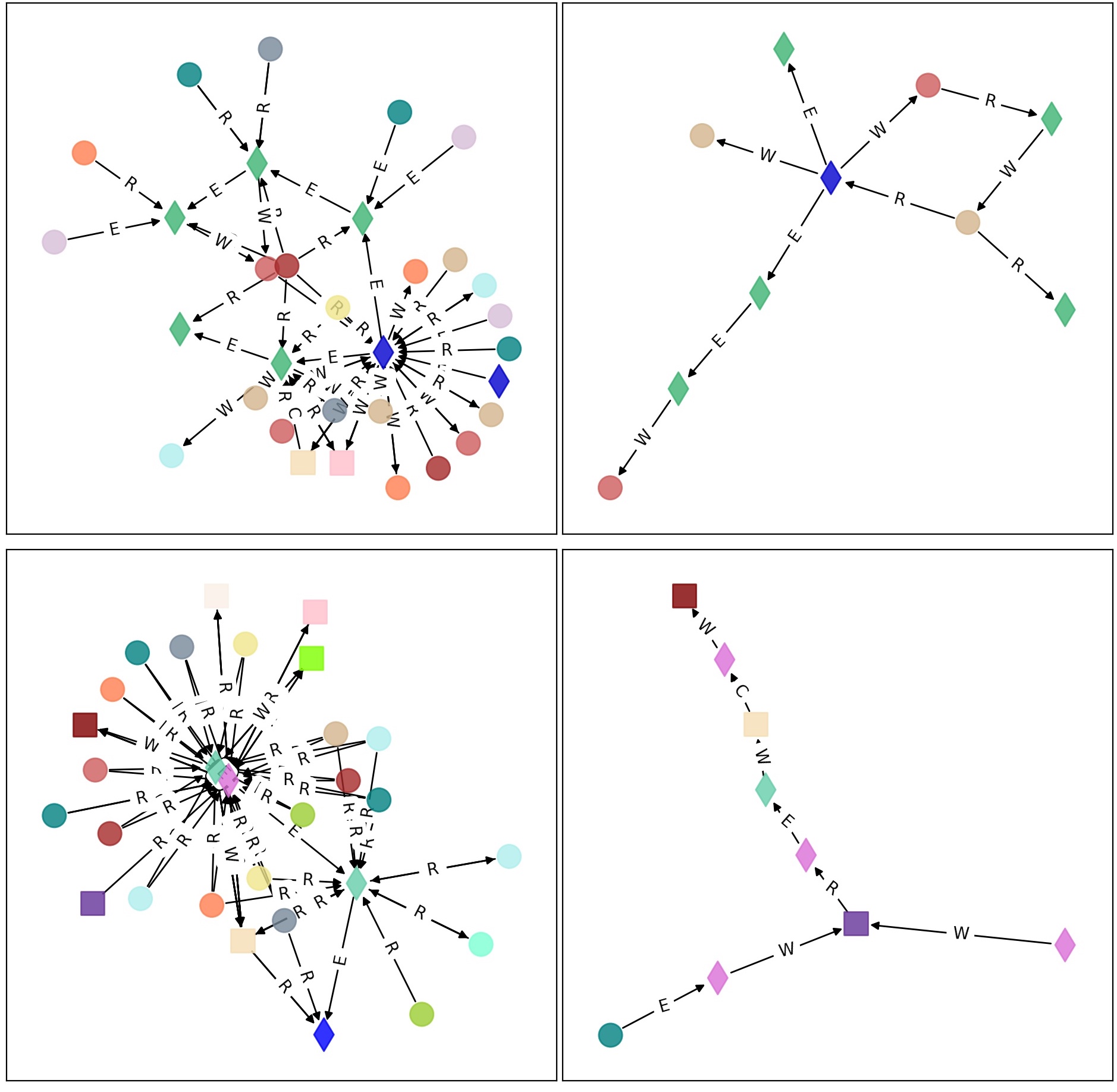}
        \centerline{(a)}
      \end{minipage}
      \begin{minipage}[a]{1\columnwidth}
        \centering
        \includegraphics[width=0.9\columnwidth]{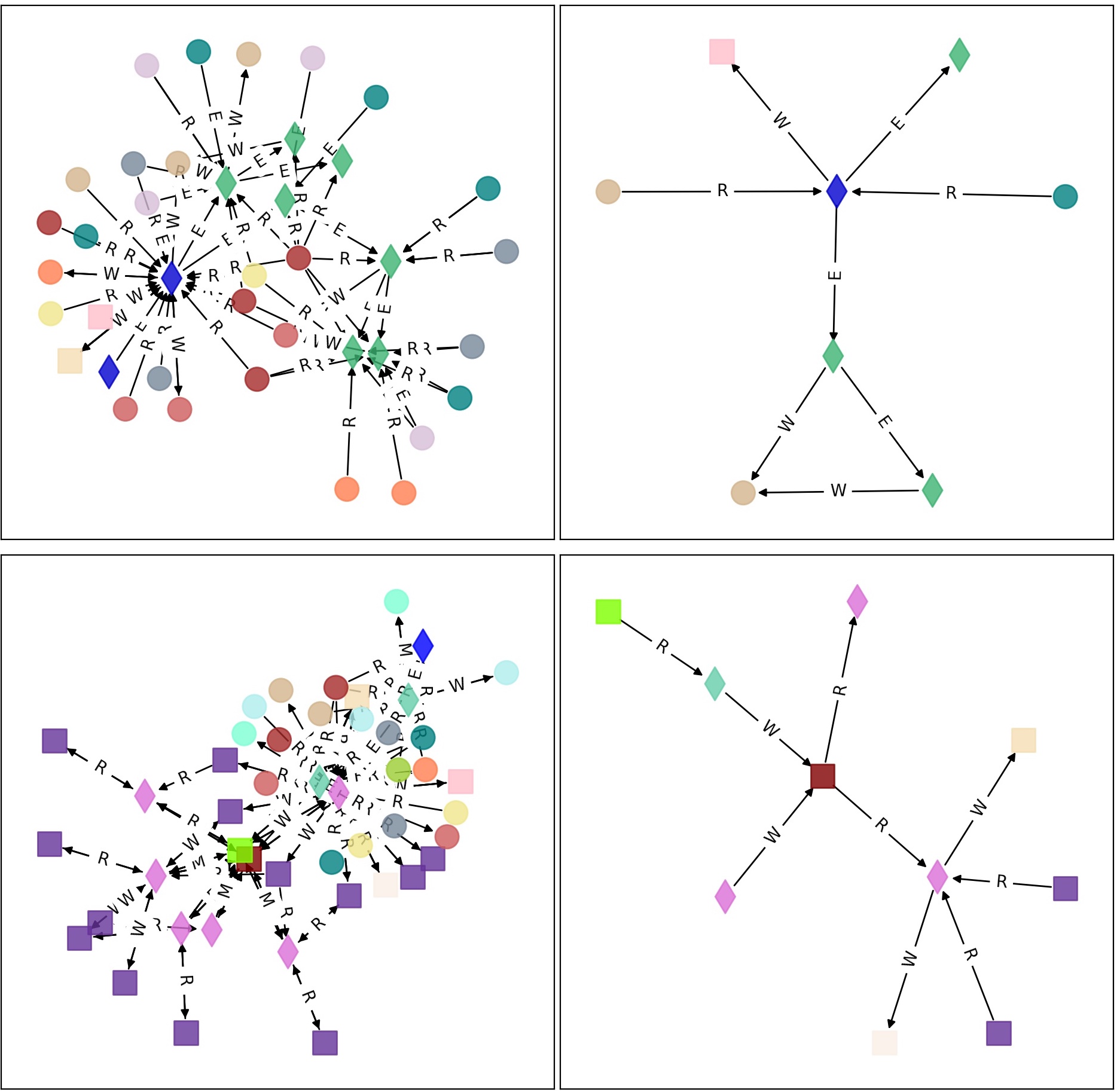}
        \centerline{(b)}
    \end{minipage}
    
\caption{Sample (a) negative and (b) positive pairs, i.e., $(\mathcal{G}^{-},\mathcal{G}_p)$ and $(\mathcal{G}^{+},\mathcal{G}_p)$ pairs defined in Sec. \ref{subsec:sample_generation}.}
  \label{fig:neg_pos_queries}
\end{figure*}

\section{Robustness Comparison}
\label{app:robust_generic}
\begin{figure}[!ht]
\centering
  \begin{minipage}[a]{0.48\columnwidth}
    \centering
    \includegraphics[width=1\columnwidth]{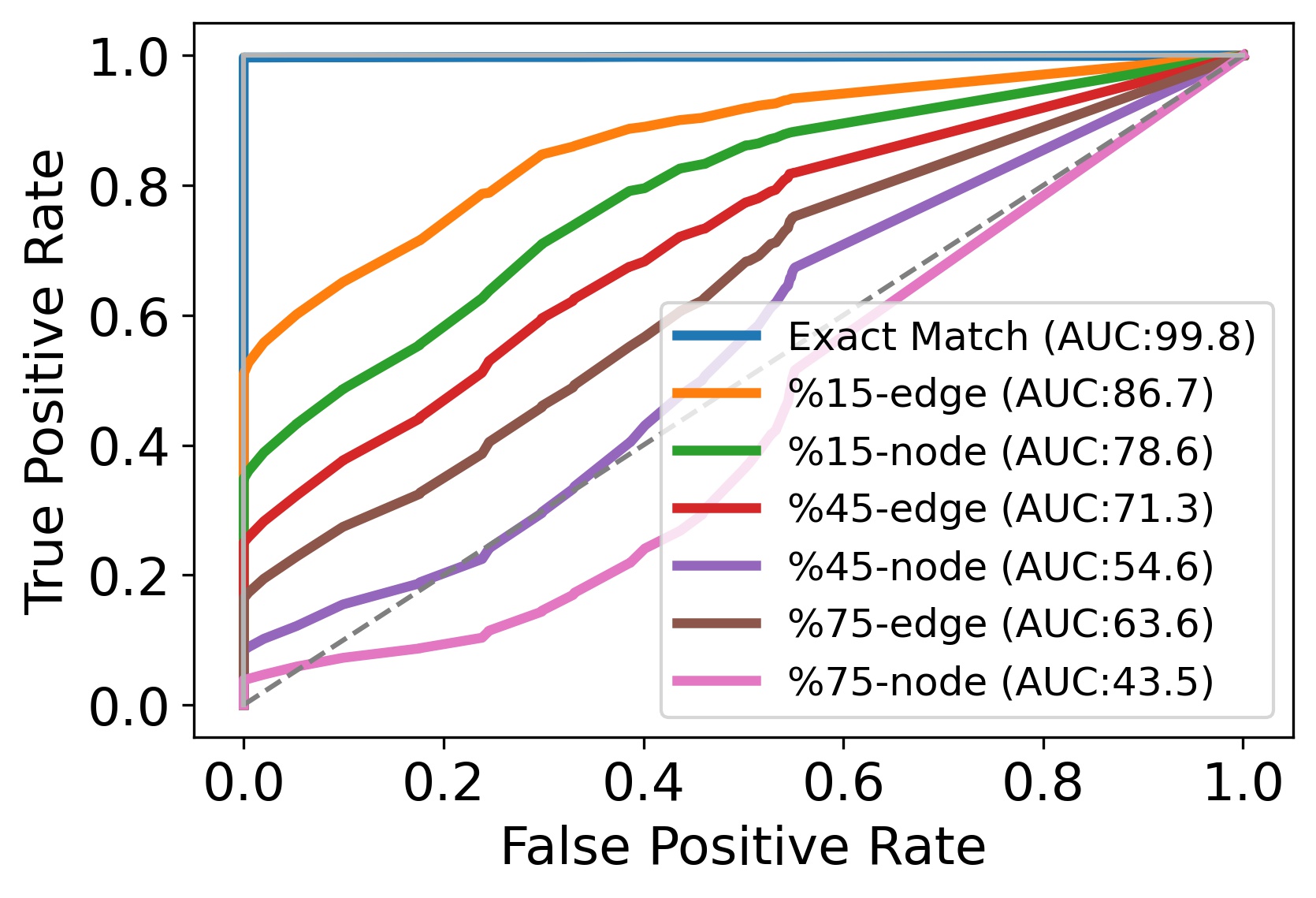}
    \centerline{(a)}
  \end{minipage}
  \begin{minipage}[a]{0.48\columnwidth}
    \centering
    \includegraphics[width=1\columnwidth]{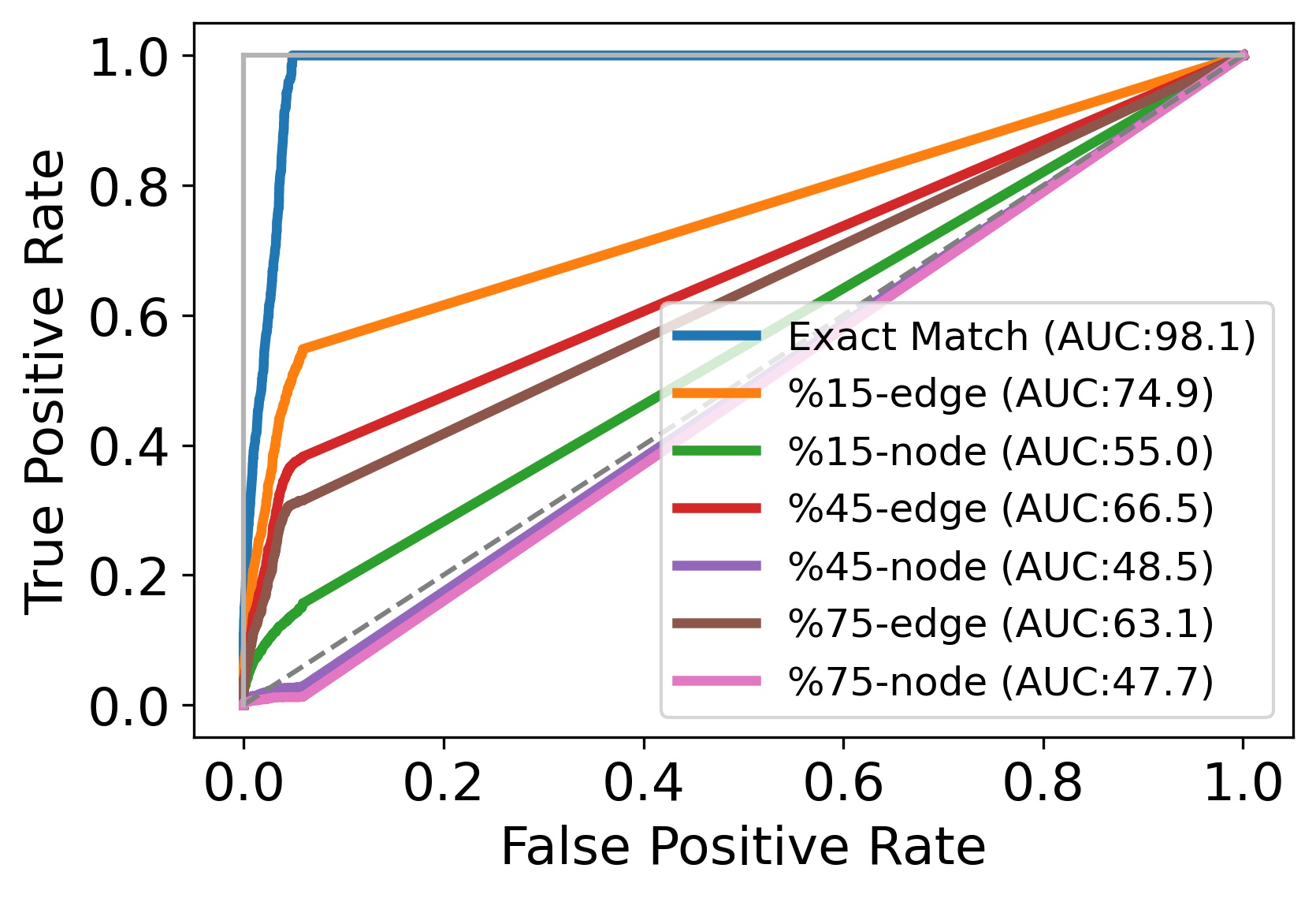}
    \centerline{(b)}
  \end{minipage}
   \caption{ROC curves for searching generic behaviors in the Theia dataset using (a) \system and (b) Poirot, considering both exact matching and imprecise matching with a portion of query edges and nodes are removed in the provenance graph.}
  \label{fig:system_roc}
\end{figure}

To examine the overall robustness of our technique in handling imprecise queries, we conducted two additional tests. 
First, we adopt the same approach described in Sec.~\ref{subsec:powerorderembedding} to randomly remove a portion of query edges and nodes from the test samples generated in Sec. \ref{sec:GED_res}.
As depicted in Fig.~\ref{fig:system_roc}(a), the removal of 15\% of edges leads to a 13\% decrease in the model's accuracy. 
This demonstrates that our subgraph matching score is sensitive to alterations in the input data, yet still maintains a relatively high level of accuracy.
We also assess the robustness of Poirot when there is a discrepancy between the queried behavior and its actual version in the provenance graph. 
As shown in Fig~\ref{fig:system_roc}(b), Poirot's node alignment performance declines significantly even with the absence of a single edge or node within the query graph.

\textcolor{black}{In the second test, we evaluate our technique's ability to identify mutations in attack behavior. To conduct this assessment, we examine three distinct malware families, each containing two samples. One of the samples within each family has its corresponding provenance graph available in \cite{satvat2021extractor}. For the second sample in each family, we utilized a dynamic malware analysis platform\footnote{https://any.run} to analyze the actions performed by the sample on the system and generate the corresponding provenance graphs.
We proceeded by testing whether the subgraph relationship between the pairs of isolated provenance graphs is preserved. 
Given that these malware families target the Windows operating system, we utilized our model trained on the Five Directions dataset for evaluation. In each scenario, we treat each pair of graphs as a query and target. These graph pairs are processed through our system, similar to the other tests, to calculate the resulting order violation penalty. This penalty is used to assess the extent of matching between the query and target graphs.
Results show that all three pairs can be matched using our pretrained model with the same parameters used in our tests. 
Each test is carried out twice by interchanging query and target graphs, resulting in the values presented in the last two columns.
The malware hashes and the resulting matching scores are presented in Table \ref{tab:malware}.
The outcomes reveal that when queried with mutated behaviors, \system can attain a matching score of 0.5 or higher.
This finding further underscores the robustness of our technique against real-world mutations in behavior.  
}

\begin{table}[!]
\centering
\caption{\textcolor{black}{Matching Scores for Mutated Malware Behaviors}}
\resizebox{\linewidth}{!}{
\begin{tabular}{|c|c|c|c|c|}
    \hline
    Malware  &Report & Malware Sample  & \multicolumn{2}{c|}{Matching } \\
    Family   &Source   & MD5    & \multicolumn{2}{c|}{Score }  \\ 
    \hline
    Carbanak &Kaspersky\cite{Carbanak} &B8E1E5B832E5947F41FD6AE6EF6D09A1 & 0.83 & 0.56 \\
    njRAT &  Fidelis\cite{Fidelis}&8B482947F8AA69E7F21BB5D51C363135 & 0.75 & 0.66  \\
    HawkEye & Fortinet\cite{HawkEye} &E20FF757A8A3E61CD78528C83D8DC796 & 0.88 & 1.00  \\
    \hline
\end{tabular}
}
\label{tab:malware}
\end{table}

\section{Sample Query Graphs}
\label{sec:app-query}
\textcolor{black}{We provide visualizations of positive and negative graph pair samples, which are utilized for training and testing our models. A negative pair consists of a target graph and a query graph that do not fulfill the subgraph relationship, as depicted in Fig. \ref{fig:neg_pos_queries}(a). Conversely, a positive graph pair adheres to the subgraph relationship, Fig. \ref{fig:neg_pos_queries}(b).}

\section{Evaluation with an Unbalanced Test Set}

\label{sec:app-grid}
\begin{table}[!]
\footnotesize
\centering
\caption{\textcolor{black}{Performance Comparison on Unbalanced Datasets}}
\resizebox{\linewidth}{!}{
\begin{tabular}{|l|l|c|c|c|c|c|c|c|c|}
    \hline
    \multirow{2}{*}{\rotatebox[origin=c]{90}{}} & \multirow{2}{*}{Method} & \multicolumn{2}{c|}{1:1} & \multicolumn{2}{c|}{1:10} & \multicolumn{2}{c|}{1:100} & \multicolumn{2}{c|}{1:1000} \\ 
    \cline{3-10} 
    & & MCC & F1 & MCC & F1 & MCC & F1 & MCC & F1\\
    \hline
    \multirow{5}{*}{\rotatebox[origin=c]{90}{Theia}} & 
    IsoRankN &31.56&62.85&16.13&19.66&5.46&2.49&1.73&0.26\\
     & SimGNN & 67.39&84.49&41.51&41.72&15.17&6.88&4.87&0.74\\
     & DeepHunter & 67.65&84.42&43.44&44.21&16.13&7.67&5.19&0.83\\
     & Poirot & 95.04&97.44&79.34&79.75&39.52&28.26&13.53&3.79\\
     & \system & \textbf{99.26}  & \textbf{99.84} & \textbf{99.40} & \textbf{99.45}  & \textbf{97.72} & \textbf{97.73} & \textbf{84.34} & \textbf{83.24} \\
    \hline
    \multirow{5}{*}{\rotatebox[origin=c]{90}{Trace}} & 
    IsoRankN & 17.14 & 36.46 & 11.43 & 20.72 & 4.09 & 3.89 & 1.31 & 0.43 \\
     & SimGNN & 53.6&78.63&30.26&31.87&10.62&4.59&3.39&0.48 \\
     & DeepHunter& 51.15&77.45&29.16&31.38&10.24&4.52&3.27&0.47\\
     & Poirot& 96.02 & 98.01 & 85.69 & 86.37 & 48.63 & 39.47 & 17.42 & 6.14 \\
     & \system & \textbf{99.18} & \textbf{99.33} & \textbf{98.97} & \textbf{99.06} & \textbf{94.18} & \textbf{94.11} & \textbf{67.39}&\textbf{62.68}\\
    \hline
    \multirow{5}{*}{\rotatebox[origin=c]{90}{Cadets}} & IsoRankN & 31.89&44.15&26.15&32.24&10.31&8.72&3.35&1.05\\
     & SimGNN & 69.74&85.55&43.68&43.73&16.11&7.43&5.18&0.80\\
     & DeepHunter & 68.98&85.21&42.98&43.09&15.81&7.25&5.08&0.78 \\
     & Poirot &96.39&98.16&94.73&95.20&75.17&73.12&34.57&22.01\\
     & \system & \textbf{99.57}&\textbf{99.76}&\textbf{99.48}&\textbf{99.61}&\textbf{97.88}&\textbf{97.89}&\textbf{84.53}&\textbf{83.41} \\
    \hline
\end{tabular}
}
\label{tab:unbalanced_table}
\end{table}

\textcolor{black}{
The comparison results presented in Table \ref{tab:combined_table} were initially obtained using a balanced set of positive and negative samples. To better simulate real-world scenarios, we conducted further experiments, as outlined in Sec. \ref{sec:comparison}, by introducing class imbalance with a focus on negative samples.
In these tests, we retained all 5K negative samples and reduced the number of positive samples to achieve an imbalance ratio of $1:k$, where $k$ takes values from the set ${10, 100, 1000}$. This means that for every $k$ negative samples tested, only one positive sample was included in the evaluation process.
To ensure rigorous evaluation, we employed k-fold cross-validation, guaranteeing the inclusion of all positive samples during testing. The average performance values are presented in Table \ref{tab:unbalanced_table} using balance-aware metrics like Matthews Correlation Coefficient (MCC) and F1-score. This comprehensive approach provides valuable insights into \systemNoSpace's performance across various class distribution scenarios.}

\end{document}